\documentclass[aps,prd,preprint]{emulateapj}
\pdfoutput=1 
\usepackage[breaklinks,colorlinks,urlcolor=blue,citecolor=blue,linkcolor=blue]{hyperref}
\usepackage{amsmath,amstext}
\usepackage[T1]{fontenc}
\usepackage{apjfonts} 
\usepackage[figure,figure*]{hypcap}
\usepackage{color, colortbl}
\usepackage[table]{xcolor}

\definecolor{Gray}{gray}{0.9}

\shorttitle{The Role of Cosmic Ray Transport in Shaping the Simulated CGM}
\shortauthors{Butsky and Quinn}

\begin{document}

\title{The Role of Cosmic Ray Transport in Shaping the Simulated Circumgalactic Medium}

\author{Iryna S. Butsky}
\author{Thomas R. Quinn}

\affiliation{Astronomy Department, University of Washington, Box 351580, Seattle, WA 98195-1580, USA}


\begin{abstract}
The majority of galactic baryons reside outside of the galactic disk in
the diffuse gas known as the circumgalactic medium (CGM). 
While state-of-the art simulations 
excel at reproducing galactic disk properties, many struggle to drive strong galactic
winds or to match the observed ionization structure of the CGM using only thermal supernova feedback. 
To remedy this, recent studies have invoked non-thermal cosmic ray (CR) 
stellar feedback prescriptions. However, numerical schemes of CR transport are still
poorly constrained.
We explore how the choice of CR transport affects the multiphase
structure of the simulated CGM.  We implement anisotropic
CR physics in the astrophysical simulation code, {\sc Enzo}
and simulate a suite of isolated disk galaxies 
with varying prescriptions for CR transport: isotropic diffusion, anisotropic diffusion, 
and streaming. We find that all three transport
mechanisms result in strong, metal-rich outflows  but differ in the
temperature and ionization structure of their CGM. 
Isotropic diffusion results in a spatially uniform, warm CGM that underpredicts
the column densities of low-ions. Anisotropic diffusion develops a reservoir of cool gas
that extends further from the galactic center, but disperses rapidly with distance. 
CR streaming projects cool gas out to radii of 200 kpc, supporting a truly multiphase medium. In addition, 
we find that streaming is less sensitive to changes in 
constant parameter values like the CR injection fraction, transport velocity, and 
resolution than diffusion. We conclude that CR streaming is a more robust implementation of CR
transport and motivate the need for detailed parameter studies of CR transport.

\end{abstract}

\section{Introduction}
The majority of baryons in galactic halos, including the majority of metals, 
reside in the circumgalactic medium (CGM) of galaxies \citep{Werk:2013, Werk:2014}. 
Loosely defined, the CGM refers to the diffuse, multiphase gas that extends
to the virial radius of galaxies. The CGM is shaped by the interplay 
between outflows from the star-forming disk and
inflows from the pristine intergalactic medium (IGM) and 
provides constraints to theories of galaxy formation and evolution. 

Early theoretical works 
predicted the existence of the CGM as a by-product of the interactions between
cooling and accretion during galaxy formation \citep{Binney:1977,Rees:1977,
Silk:1977}. Due to its low density, the CGM is extremely difficult to 
observe directly in emission. Instead, recent observations, such as 
those taken with The Cosmic Origins Spectrograph (COS) \citep{Green:2012}
on board the Hubble Space Telescope (HST), have studied the CGM through
absorption lines in quasar spectra that intersect the halos of galaxies
along the line-of-sight.
Using the quasar absorption method, different groups have greatly
advanced our knowledge of the CGM both 
in high-redshift (e.g. \citealt{Steidel:2010,Rudie:2012,Turner:2014}), 
and low-redshift (e.g. \citealt{Chen:2010,Gauthier:2010,Kacprzak:2010,Bordoloi:2011, 
Prochaska:2011, Tumlinson:2011, Nielsen:2013, Stocke:2013, Werk:2013, Zhang:2016}) galaxies.
Today, we understand the CGM to have an intricate and complex 
temperature, density, and kinematic structure that is shaped
by galactic outflows. 

Simulations play an integral role in understanding the physical processes 
that govern galactic outflows by allowing astronomers to perform 
experiments testing the validity of different feedback prescriptions. 
The success of a simulation has traditionally been marked
by its ability to reproduce observed galactic disk properties,
such as the morphology, Tully-Fisher relation,
and star formation rate density \citep{Schaye:2010, Dave:2011, 
Puchwein:2013, Stinson:2013, Christensen:2014}. Requiring that simulations
also match the observed structure of the CGM
will place strong additional constraints to stellar feedback models. 

One such constraint is that stellar feedback must drive galactic
outflows that carry a substantial amount of metals along with the gas they expel. 
Although metals are produced within galactic disks, galaxies retain only 
$\sim 20-25\%$ of these metals in their stars and ISM \citep{Peeples:2014}.
Data from the Sloan Digital Sky Survey suggest that metals have been lost to outflows
\citep{Tremonti:2004}. Since galaxies have very low metallicities at early cosmic 
times ($z > 3$), the metals ejected by supernovae serve as excellent tracers of
outflowing material and can be used to make predictions for future observations.

These outflows must not only enrich the CGM but also reproduce its
multiphase ionization structure. For example, the CGM of galaxies at low redshift 
contains a substantial amount of metal-enriched, cool gas at $10^4-10^5$K 
\citep{Werk:2013,Werk:2014}. However, cooling times of $\simeq 10^5$K gas are 
very short compared to galactic timescales, and it is unclear how this
material survives in such abundance. The data seem to imply an additional
unknown source of non-thermal pressure that supports the warm gas against 
condensation.

Recent studies have investigated various types of thermal wind-launching 
mechanisms, including radiation pressure from massive stars 
\citep{Kim:2011, Murray:2011,Hopkins:2012,Sharma:2012,Wise:2012},
thermal supernova feedback \citep{Abadi:2003, Joung:2009, Hummels:2012, 
Creasey:2013}, kinetic supernova feedback \citep{Hopkins:2012,Springel:2003,
Agertz:2013}, supernova superbubble models, 
\citep{Keller:2015, Keller:2016},  and subgrid models tuned to generate
strong outflows \citep[e.g.][]{Springel:2003, Stinson:2006, Oppenheimer:2006, Governato:2012}).
However, many of these feedback prescriptions still struggle to produce sufficiently
strong galactic outflows.
Those that do succeed in expelling gas from the galactic
disk underpredict the observed column densities of H I, O VI, and low ions 
in the CGM.
\citep{Hummels:2013, Marasco:2015,Fielding:2017,Gutcke:2017}. 
This discrepancy is amplified in massive galaxies for which thermal supernova feedback 
also struggles to quench star formation \citep{Pontzen:2013}. In such
galaxies, feedback from active galactic nuclei (AGN) is often invoked to regulate
the baryon cycling process 
\citep[e.g.][]{Suresh:2015,Tremmel:2017, Oppenheimer:2017}. 
Although AGN help recreate some observable properties 
for massive galaxies, they cannot account for discrepancies in galaxies (such as 
the Milky Way) that host a small supermassive black hole (SMBH) at their core.
It is likely that the missing key is a non-thermal component.

Cosmic rays (CRs) are charged particles that have been
accelerated to relativistic speeds in shocks (e.g.\ SNe \citealt{Ackermann:2013},
structure formation shocks \citealt{Pfrommer:2008} radio-loud AGN
\citealt{McNamara:2007}) and are observed to be roughly in
equilibrium with the thermal and magnetic pressures in our galaxy \citep{Boulares:1990}.
Due to past computational constraints, CRs have only recently been included in 3-D
hydrodynamical simulations of galactic structure. These simulations have shown that
CR feedback drives strong, mass-loaded  outflows and suppresses star formation 
\citep[e.g.][]{Miniati:2001, Ensslin:2007,Jubelgas:2008, Socrates:2008, Uhlig:2012, Vazza:2012, 
Booth:2013, Salem:2014a, Girichidis:2016, Simpson:2016, Samui:2017}.
Furthermore, CRs provide pressure support to the thermal gas, which may explain the
presence of the observed structures in the CGM that appear to be out of thermal equilibrium.

 \cite{Salem:2016} were the first to show that stellar feedback models that 
included CR  energy were better
at matching COS-Halos data for low-ion column densities
than those with purely thermal feedback. However, these results depended
on the choice of a constant diffusion coefficient, which is only loosely constrained
by observations. Furthermore, their simulations neglected magnetic fields, which
are crucial for accurate modeling of CR transport. 
Traditionally, implementations of CR transport have been separated into two
approximations: diffusion and streaming. Although both approaches have been
successful at driving galactic outflows, the strength and mass-loading
factor of CR driven winds depends on the invoked
transport mechanism \citep{Ruszkowski:2017,Wiener:2017}. 

In this work, we investigate the role of different CR transport prescriptions
in shaping the multiphase structure of the CGM. 
This paper is structured as follows. In \S \ref{sec:crenzo}, we describe
the implementation of CR physics in {\it ENZO}. In \S \ref{sec:methods}, 
we describe the simulation suite and the relevant initial conditions and physical
modules used in our isolated disk setup. We present our results in \S
\ref{sec:results}, focusing on the generated outflows, temperature structure, and
column densities of the different galaxy models. 
We outline the qualitative differences between CR diffusion and streaming and discuss
future prospects in \S \ref{sec:discussion}. Finally, we provide a summary of 
our work in \S \ref{sec:summary}. 


\begin{table*}[ht]
\begin{minipage}{180mm}
\begin{center}
\caption{Simulation Parameters }
\footnote{A summary of the simulation initial conditions. 
The run ID and the corresponding boldface entries describe the
fiducial runs discussed in depth throughout this paper. The
fiducial runs differ from each other only by the invoked CR
transport mechanism. We deviate from the fiducial runs by 
changing the resolution, the fraction of supernova energy 
injected as CRs ($\mathrm{f_c}$), the diffusion coefficient 
($\kappa_{\varepsilon}$), the streaming factor ($f_s$), and
including the CR heating term ($H_c$). Figure \ref{fig:CRBeta}
compares the effect of these variables on the CR pressure 
distribution in the CGM. 
}
\begin{tabular}{l || c | ccccc}
\hline \hline
Run ID & Min. grid size (pc) & \multicolumn{5}{c}{CR parameters} \\
\cline{3-7}
& & $\mathrm{f}_{c}$  & Transport Mode & $\kappa_{\varepsilon}$ [$\times 10^{28}$ cm$^2$/s] & $f_s$ & $H_c$ \\
\hline 
\bf {\sc ncr}    & \bf 160 &\bf  -  & \bf no CRs           & \bf - &\bf  - & \bf no \\ 
		    & 320 & -  & no CRs           & - & -  & no \\ 
\\
{\sc adv}  & \bf 160 &  \bf 0.1  & \bf advection only  & \bf 0 & \bf - & \bf no \\ 
           & 320 &  0.01  & advection only  & 0 & -  & no \\ 
           & 320 &  0.1 & advection only  & 0 & -  & no \\ 
           & 320 &  0.3 & advection only  & 0 & -  & no \\ 
\\
{\sc isod}   & \bf 160 &  \bf 0.1  & \bf isotropic \bf diffusion  & \bf 3 & \bf -  & \bf no \\ 
			 & 320 &  0.01 & isotropic diffusion  & 3 & - & no  \\ 
			 & 320 &  0.1  & isotropic diffusion  & 3 & - & no  \\ 
			 & 320 &  0.3  & isotropic diffusion  & 3 & - & no  \\ 
             & 320 &  0.1  & isotropic diffusion  & 1 & - & no  \\ 

\\
{\sc anisd} & \bf 160 &   \bf 0.1  & \bf anisotropic diffusion & \bf 3 & \bf - & \bf no  \\ 
            & 320 &   0.01  & anisotropic diffusion & 3 & -  & no \\ 
 		   & 320 &   0.1  & anisotropic diffusion & 3 & - & no  \\ 
		   &320 &    0.1  & anisotropic diffusion & 10 & - & no \\ 
 		   & 320 &   0.3  & anisotropic diffusion & 3 & - & no  \\ 
 		   & 320 &   0.1  & anisotropic diffusion & 1 & - & no  \\ 
\\
{\sc anisdh} & \bf 160 &   \bf 0.1  & \bf anisotropic diffusion & \bf3 &\bf  - & \bf yes   \\
            	   & 320 &   0.1  & anisotropic diffusion & 3 & - & yes  \\ 
\\

 {\sc stream}&\bf 160 & \bf  0.1 & \bf  streaming & \bf  - & \bf  4 &\bf yes \\ 
             & 160 &   0.1  &  streaming & - & 1 & yes  \\ 
             & 320 &   0.01  &  streaming & - & 4  & yes \\ 
             & 320 &   0.1  &  streaming & - & 1  & yes \\ 
             & 320 &   0.1  &  streaming & - & 2  & yes \\ 
             & 320 &   0.1  &  streaming & - & 4 & yes \\ 
             & 320 &   0.3  &  streaming & - & 4  & yes \\

\hline  
\end{tabular}
\end{center}
\end{minipage}
\label{tab:parameters}
\end{table*}

\section{Cosmic Rays in ENZO} \label{sec:crenzo}
In the following section, we describe our implementation of the CR 
fluid into the different Riemann solvers in the adaptive mesh refinement 
(AMR) magnetohydrodynamics (MHD) simulation code {\sc Enzo}.  
Our implementation builds on the work of \cite{Salem:2014a}, which described the
integration of a CR fluid in the {\sc zeus} finite-difference solver  \citep{Stone:1992}.  
Because the current implementation of the {\sc zeus}
solver in the public version of {\sc Enzo} doesn't support MHD, the 
primary advantage of our work is the ability to model the 
interaction of CRs with magnetic field lines. In  \S \ref{sec:crfluid},  
we describe the new set of conservation equations. In \S \ref{sec:crstreaming} and \S \ref{sec:crdiffusion}, 
we describe the algorithm for anisotropic CR streaming and diffusion.
In \S \ref{sec:gradient} we discuss our approach for avoiding unphysical values
of CR energy. 


\subsection{Cosmic Ray Fluid}\label{sec:crfluid}
As an Eulerian code, {\sc Enzo} models gas as a fluid moving through
grid cells that are fixed in space. 
At every timestep, {\sc Enzo} advances the state of the fluid in the 
simulation by numerically approximating a solution to the equations below. 
These equations encompass the conservation of mass (Equation \ref{eqn:mass}), 
conservation of momentum (Equation \ref{eqn:momentum}), the induction equation
(Equation \ref{eqn:induction}), and the energy equation (Equation 
\ref{eqn:energy}). In simplified terms, these equations assert that the change
over time in the value of a given conserved quantity in one cell 
(the $\frac{\partial }{\partial t}$ term ) is equal to the flux of that conserved 
quantity through the boundaries of that cell (the $\nabla \cdot ()$ term). Source terms, 
which encompass both energy gains and losses (for example, an injection of CR energy
after a supernova explosion) appear on the right-hand-side of the equality. Traditionally, the
evolution of thermal gas is encompassed in Equations \ref{eqn:mass} - \ref{eqn:energy}
(setting the cosmic ray pressure term, $P_c$ to zero). 
We model the evolution of CRs as an additional ultra-relativistic proton fluid with adiabatic
index $\gamma_{c} = 4/3$ \citep{Jun:1994, Drury:1986} in Equation \ref{eqn:crenergy}.

\begin{equation}\label{eqn:mass}
\frac{\partial \rho}{\partial t} + \nabla \cdot (\rho {\bf v}) = 0
\end{equation}

\begin{equation}\label{eqn:momentum}
\frac{\partial(\rho {\bf v})}{\partial t} + \nabla 
	\cdot (\rho{\bf vv}^T + P_g + P_c) = -\rho \nabla {\bf \Phi}
\end{equation}

\begin{equation}\label{eqn:induction}
\frac{\partial \bf B}{\partial t} + \nabla\cdot({\bf Bv^{\mathrm{T}}
- vB^{\mathrm{T}}}) = \bf 0
\end{equation}

\begin{equation}\label{eqn:energy}
\frac{\partial \varepsilon_g}{\partial t} + \nabla\cdot ({\bf v} \varepsilon_g)
= - P_g\nabla\cdot{\bf v} + H_c + 
\Gamma_g + \Lambda_g
\end{equation}

\begin{equation}\label{eqn:crenergy}
\frac{\partial \varepsilon_c}{\partial t} + \nabla\cdot {\bf F}_\mathrm{c}
= - P_\mathrm{c} \nabla\cdot{\bf v} - H_\mathrm{c} + 
\Gamma_\mathrm{c} + \Lambda_\mathrm{c}.
\end{equation}
\begin{equation}\label{eqn:crtransport}
{\bf F}_\mathrm{c} = {\bf v} \varepsilon_{\mathrm{c}} +{\bf v}_\mathrm{s}(\varepsilon_{\mathrm{c}}+P_{\mathrm{c}})-\kappa_{\varepsilon}{\bf b}({\bf b\cdot\nabla}
\varepsilon_{\mathrm{c}})
\end{equation}
Here, we define $\Phi$ to be the gravitational potential 
 (where $ \bf \nabla ^2 \Phi = 4\pi G \rho_{tot} $), $\rho_{tot}$ to be the total density, 
and $\rho$, $\bf v$ to be the gas density and velocity. 
$\bf B$ is the magnetic field strength, and $\bf b = 
{\bf B}/|{\bf B}|$ is the magnetic field direction. 
The superscript $\mathrm{T}$ denotes the vector transpose.
The internal gas pressure, $P_g$ is related to the internal thermal energy 
density $\varepsilon_g = (\gamma_g - 1)P_g$, where $\gamma_g = 5/3$. 
Similarly, the CR pressure is related to the CR energy density, 
$\varepsilon_c = (\gamma_c - 1)P_c$, with $\gamma_c = 4/3$. 
From here on out, we use the subscript 'g' to refer to properties of the
thermal gas, and the subscript 'c' to refer to properties of the CR fluid. 
We model the diffusion coefficient, $\kappa_{\varepsilon}$ as a constant, which is
observationally constrained to be on the order of $\kappa_{\varepsilon} \simeq
10^{28}$ cm$^2$ s$^{-1}$ \citep{Ptuskin:2006, Strong:1998, Tabatabaei:2013}. 
Although we neglect CR transport perpendicular to the magnetic field, 
it could have observable effects \citep{Kumar:2014}. 
$\Gamma$ and $\Lambda$ are energy source and sink terms respectively. 
In our simulations, the source of CR energy is supernova events. We do not 
implement hadronic CR energy losses. 

Equation \ref{eqn:crtransport} encompasses the three modes of CR transport that
we have implemented. The first term on the left represents advection, in which
the CR fluid moves with the bulk velocity of the thermal gas. This term is solved explicitly 
in the Riemann solvers of {\sc Enzo}. The next two terms describe CR streaming and diffusion respectively. 
These are approximations to CR motion relative to the gas and are therefore solved 
 separately from the advection equation. The implementation of these transport methods 
is described in detail in sections \ref{sec:crstreaming} and \ref{sec:crdiffusion}.  

In the streaming approximation, CRs move relative to the gas
with a velocity given by

\begin{equation}
{\bf v}_{\mathrm{s}} = -sgn({\bf b} \cdot {\bf \nabla}\varepsilon_{\mathrm{c}}).
f_s{\bf v_A}
\end{equation}
Here, $f_s$ is the constant streaming factor described in \citet{Ruszkowski:2017}. 
The  Alfv{\`e}n velocity, ${\bf v_A} = {\bf B}/\sqrt{4\pi\rho}$ represents the transverse waves
propagating along the magnetic field lines in a plasma.
The function $sgn$ returns the sign of the enclosed
expression.
In this limit, CRs also transfer momentum to 
the thermal gas through the heating term, $H_c$, where
\begin{equation}
H_c = |{\bf v_A}\cdot \nabla P_{c} |.\end{equation} 
Although this term appears
only in the streaming approximation, we follow the example of \citet{Wiener:2017} and
include it in some simulations with CR diffusion to isolate the underlying differences
in the different transport mechanisms (see Table \ref{tab:parameters}). 

\subsection{Cosmic Ray Streaming}\label{sec:crstreaming}
Both CR streaming and diffusion are approximations to CRs scattering off of 
Alfv{\`e}n waves. 
The difference lies in the source that is generating these waves. 
In the streaming approximation, CRs are assumed to drive the growth
of Alfv{\`e}n waves through the streaming instability \citet{Kulsrud:1969}. 
This is often referred to as the "self-confinement" case \citep{Zweibel:2017}.

We isolate the streaming behavior from the general CR transport equation (Eqn. \ref{eqn:crtransport}) using
\begin{equation}
\frac{\partial \varepsilon_c}{\partial t} - \nabla \cdot [{\bf v_{\mathrm{s}}} 
(\varepsilon_{\mathrm{c}}+P_{\mathrm{c}})] = 0.
\end{equation}
At each simulation timestep, we calculate CR streaming by updating
the value of the CR energy density in each cell. The evolution of the
CR energy density in cell $i$ is given by
\begin{equation}\label{eqn:crstreaming}
\varepsilon_{c, i}^{n+1} = \varepsilon_{c, i}^{n} -\Delta t \sum_j 
-{\bf \mathrm{sgn}}({\bf b}_{ij} \cdot {\bf \nabla}\varepsilon^n_{\mathrm{c},ij}){\bf v}_{A,ij}
\cdot {\bf n}_{ij} (\Delta x_{ij})^{-1}, 
\end{equation}
where $\varepsilon^n_{c,i}$ is the value of the CR energy density in cell $i$ before 
streaming is applied. $\Delta t$ is the timestep, and
the terms ${\bf b}_{ij}$ and $\nabla \varepsilon_{c, ij}$ 
are the direction of the magnetic field and the gradient of CR energy density 
computed at cell face $j$. 
${\bf n}_{ij}$ describes the plane parallel to face $j$, and $\Delta x_{ij}$ is the length
of the cell's axis that is perpendicular to face $j$. If the cells in the simulation
are constructed as cubes, then $\Delta x$ can be treated as a constant and moved out of
the summation term.

The streaming time step is set by the bulk motion of the gas
and the alfv{\`e}n velocity, so that
$ t_{\mathrm{stream}} < \frac{\Delta x}{{\bf v}_g + f_s{\bf v}_a}$.
We avoid instabilities near local extrema of CR energy density 
by employing the regularization described in \citet{Sharma:2009} and \citet{Ruszkowski:2017} where
\begin{equation}
{\bf F}_c = -(e_c + p_c) {\bf u}_A \mathrm{tanh} (h_c \hat{\bf v} \cdot \nabla e_c / e_c).
\end{equation}
We follow the recommendation of \citet{Ruszkowski:2017} and set the free regularization
parameter, $h_c$, to 10 kpc.


\subsection{Cosmic Ray Diffusion}\label{sec:crdiffusion}
In the "extrinsic turbulence" case, Alfv{\`e}n waves are excited by
turbulent motions in the thermal gas. Since the
gyroradius of CRs around magnetic field lines is significantly smaller than the best-resolved cells in our 
simulation, CR transport in this regime is modeled as diffusion parallel to the direction of 
magnetic field lines. At each timestep, in addition to solving the conservation 
equations, {\sc Enzo} computes CR diffusion according to
\begin{equation}
\frac{\partial \varepsilon_c}{\partial t} - \nabla \cdot [\kappa_{\varepsilon}
{\bf b} ({\bf b} \cdot \nabla \varepsilon_c )] = 0.
\end{equation}
With diffusion, the updated CR energy density in a grid cell, $i$ follows the prescription  
\begin{equation}\label{eqn:crdiffusion}
\varepsilon_{c, i}^{n+1} = \varepsilon_{c, i}^{n} + \Delta t
\sum_j \kappa_{\varepsilon} ({\bf b}_{ij} \cdot \nabla \varepsilon_{c, ij}^n)
{\bf b}_{ij} \cdot {\bf n}_{ij} (\Delta x_{ij})^{-1}, 
\end{equation}
where $\Delta t$ is the diffusion timestep.

In the case of isotropic diffusion, Equation \ref{eqn:crdiffusion} becomes

\begin{equation}\label{eqn:isocrdiffusion}
\varepsilon_{c, i}^{n+1} = \varepsilon_{c, i}^{n} + \Delta t
\sum_j \kappa_{\varepsilon}\nabla \varepsilon_{c, ij}^n
\cdot {\bf n}_{ij} (\Delta x_{ij})^{-1}.
\end{equation}

We employ an explicit diffusion scheme because we find that our simulation
time step is not limited by the diffusion time step constraint of
$t_{\mathrm{diff}} < \frac{1}{2N}\frac{\Delta x^2}{\kappa_{\varepsilon}}$, where N is 
the dimensionality of our simulation. For a detailed discussion on numerically 
modeling semi-implicit anisotropic CR diffusion, see \citet{Pakmor:2016b}.


\subsection{Limiting the Gradient}\label{sec:gradient}
For some geometrical configurations, using a simple estimate of the CR energy
gradient when calculating anisotropic diffusion violates
the second law of thermodynamics and leads to an unphysical flow of CR energy 
from cells with lower energy density to cells with higher energy density. In more extreme cases, 
this leads to some cells developing unphysical (negative) values of CR energy. 
To combat this issue, we employ a limited gradient as described in \citet{VanLeer:1977}. 

The simple gradient only considers the flux through a given cell face, so that
\begin{equation}
\nabla \varepsilon_{c} = \frac{\varepsilon_{c, i+1} - \varepsilon_{c, i}}{\Delta x_i}
\end{equation}
where $i$ and $i+1$ are the indices of neighboring cells. 

Where the simple gradient produces an unphysical flux, we estimate the limited CR energy
density gradient on the interface \citep{Sharma:2007, Pakmor:2016b} to be 
\begin{equation}
\nabla \varepsilon_{c} = \frac{4}{\sum_n (\nabla \varepsilon_{cr}^n)^{-1}}. 
\end{equation}

This approximation takes into account the component of the gradient that
is parallel to the cell face, $n$.

\begin{figure*}
\begin{center}
\includegraphics[width=\textwidth]{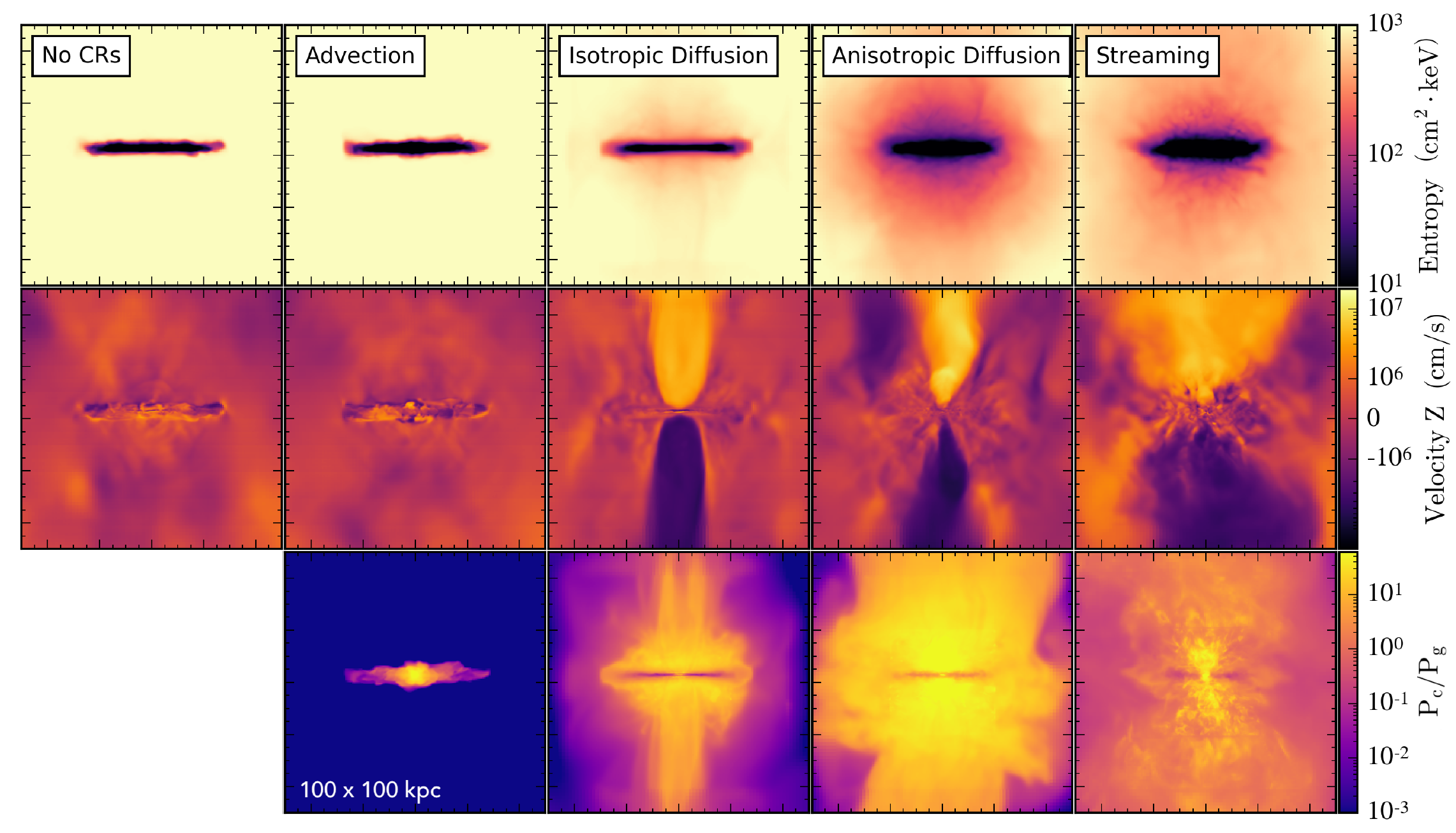}
\caption{\label{fig:outflows}
A comparison of the outflows between different galaxy models. Each row
shows the edge-on, density-weighted projection of the
gas entropy (top), z-component of the velocity (middle) and
ratio of CR and thermal pressures (bottom).
Each column holds a simulated galaxy with different CR treatment.
From left to right, the columns show simulations with: no CRs, CR advection only, 
isotropic diffusion, anisotropic diffusion, and streaming.
The snapshots were taken at t = 2 Gyr. Only models with CR transport relative
to the gas drive strong outflows by providing a non-thermal source of pressure
to lift low-entropy gas out of the disk. The strength and morphology of the outflows
is sensitive to the choice of CR transport.}
\end{center}
\end{figure*}

\section{Numerical Methods}\label{sec:methods}
We conduct our work using {\sc Enzo}, an open source multi-physics 
MHD astrophysical simulation code that employs AMR to resolve areas of interest 
\citep{Collins:2010, Bryan:2014}. At each time step, {\sc Enzo} solves the
Riemann conservation equations. We employ the local Lax-Friedrichs Riemann
solver (LLF; \citealt{Kurganov:2000}) to compute the flux at cell interfaces
and spatial reconstruction is performed with the piecewise linear method
(PLM; \citealt{VanLeer:1977}). Time stepping is carried out with the total variation
diminishing (TVD) 2nd order Runge-Kutta (RK) scheme \citep{Shu:1988}.
To avoid creating unphysical magnetic monopoles, we employ a hyperbolic
divergence cleaning approach first described by \citet{Dedner:2002}. Interested
readers should see \citet{Wang:2008,Wang:2009} for detailed discussions and 
extensive testing of the MHD formulation in {\sc Enzo}. 


\subsection{Stellar Feedback} \label{sec:supernova}
Star formation in our simulations follows the prescription described in
\citet{Cen:1992} with minor modifications. 
Stars particles are only formed in grids at the maximum level of refinement
that meet predetermined density, mass, and minimum dynamical time thresholds:
$\rho_{\mathrm{cell}} > \rho_{\mathrm{thres}}$, $M_{\mathrm{cell}} > 
M_{\mathrm{thres}}$, $t_{\mathrm{cool, cell}} < t_{\mathrm{dyn, min}}$. 
The exact values for $\rho_{\mathrm{thres}}, M_{\mathrm{thres}}, t_{\mathrm{dyn, min}}$
depend on the resolution of the simulation, because a very resolved cell may not be
capable of holding sufficient mass to satisfy the formation criteria. In our 
simulations, we choose 
$\rho_{\mathrm{thres}} = 3.0 \times 10^{-26}$ g cm$^{-3}$, 
$M_{\mathrm{thres}} = 3.0 \times 10^5 M_{\odot}$, 
$t_{\mathrm{dyn, min}} = 10$ Myr. 
Additionally, the gas in the parent cell must be collapsing
(determined by measuring negative velocity divergence) and Jeans unstable.
If all conditions are met, the
parent cell produces a star cluster particle with 10\% mass efficiency so that
the initial mass is given by 
$M^{\mathrm{init}}_{\mathrm{SC}} = 0.1 \rho_g \Delta x^3$. 

Over the course of 
120 Myr, 25\% of that mass is ejected into the
cell in which the star cluster particle resides, modeling the effects of Type II 
supernovae.  We inject $10^{51}$ ergs of energy for every $42 M_{\odot}$,
so that a star cluster
particle of $M = 3.0\times 10^{5} M_{\odot}$ expels a total of 
$E_{tot} = 5.4 \times 10^{54}$ ergs of energy. In our model, this total 
energy is comprised of
thermal, magnetic, and CR energy so that $E_{th} = f_{th}E_{tot}$,
$E_{B} = f_{B}E_{tot}$, and 
$E_{cr} = f_{cr}E_{tot}$,
where $f_{th} + f_{B} + f_{cr} = 1$.  We adopt a conservative estimate of 
$f_{B} = 0.01$, $f_{cr}  = 0.1$ \citep{Wefel:1987, Ellison:2010} and 
assume 2\% of the ejecta to be metals.


\subsection{Chemistry and Cooling}\label{chemistry}

In our simulations, we explicitly track all ion species of hydrogen. 
All other species are tracked together in a metallicity variable.
We calculate cooling using the GRACKLE chemistry library 
\citep{Bryan:2014,Smith:2017} that is integrated with {\sc Enzo}. It uses
pre-computed tables of metal cooling rates as functions of the gas density 
and temperature generated by CLOUDY \citep{Ferland:2013}. In addition, we 
consider uniform photoelectric heating of $8.5\times10^{-26}$ ergs s$^{-1}$
cm$^{-3}$ without self\-shielding \citep{Tasker:2008}.
In our idealized isolated disk setup, we do not consider the ultraviolet 
background radiation from distant quasars and galaxies. 

\subsection{Synthetic Observations}\label{sec:trident}

We use {\em Trident} \citep{Hummels:2017}, a python based tool integrated with 
yt \citep{Turk:2012} to construct ion densities and generate synthetic spectra
from our simulated datasets. For ions that are not explicitly 
tracked by the simulation, we can determine their number densities, $n_{X}$, 
in post processing with
\begin{equation}
n_{X} = n_H Z \bigg(\frac{n_X}{n_H}\bigg)_{\odot},
\end{equation}
where $n_H$ is the total hydrogen number density, Z is the metallicity (a value that
is kept track of throughout the simulation), and $(n_X/n_H)$ is the solar number 
abundance for any element that is not explicitly tracked.

Trident computes relative ion abundances in a simulation cell by considering both 
photoionization from an extragalactic UV background \citep{Haardt:2012}, and  
collisional ionization within the gas.  The collisional ionization rate is determined
by the cell's temperature, density, and metallicity, using an extensive set
of lookup tables precalculated by CLOUDY, which assume ionization abundances
as predicted by collisional ionization equilibrium (CIE). 
CIE holds even when CR pressure dominates over
thermal pressure because the underlying assumption of our model 
is that the CR fluid is collisionless. 
Any deviations from CIE (which is more likely in the dense regions of the galactic disk than in the CGM)
would likely increase of the ionization rate in low temperature gas \citep{Oppenheimer:2013}.

We use this functionality in {\em Trident} to plot column densities of different ions as a function
of impact parameter in Figure \ref{fig:col_density}.  We calculate the column densities
by defining sight-lines through the simulation box. The ion number
density is then calculated in each length element, $dl$, along this projected ray. 
The column density along the  line-of-sight is then given by the summation 
$n_{\mathrm{col}} = \sum dl \cdot n$.

\begin{figure*}
\begin{center}
\includegraphics[width=\textwidth]{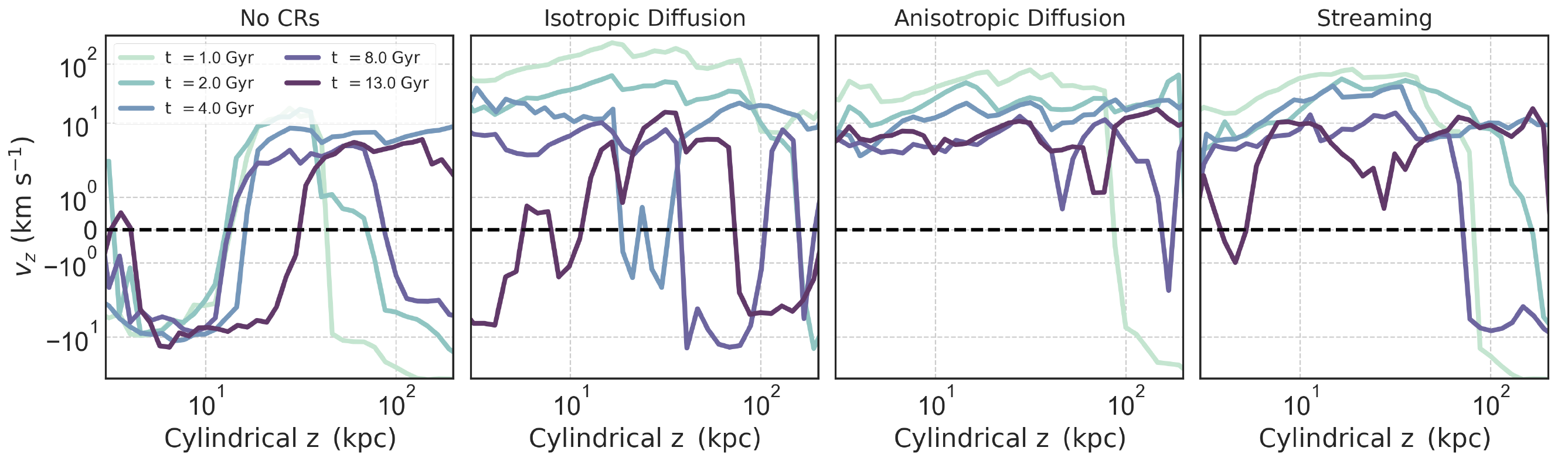}
\caption{\label{fig:velz}
The time-evolution of the vertical velocity profile as a function of the height above the galactic disk for
galaxy models {\sc ncr}, {\sc isod}, {\sc anisd}, and {\sc stream}.  Model {\sc ncr} shows no  signs of strong outflows. 
Model {\sc isod} has the fastest outflow velocities but loses its galactic wind by t = 4  Gyr. Models {\sc anisd} drive
weaker yet steadier galactic winds.  Model {\sc stream} is the only model to develop inflows near the galactic disk
while simultaneously hosting outflows at larger radii. } 
\end{center}
\end{figure*}

\begin{figure}
\begin{center}
\includegraphics[width=0.45\textwidth]{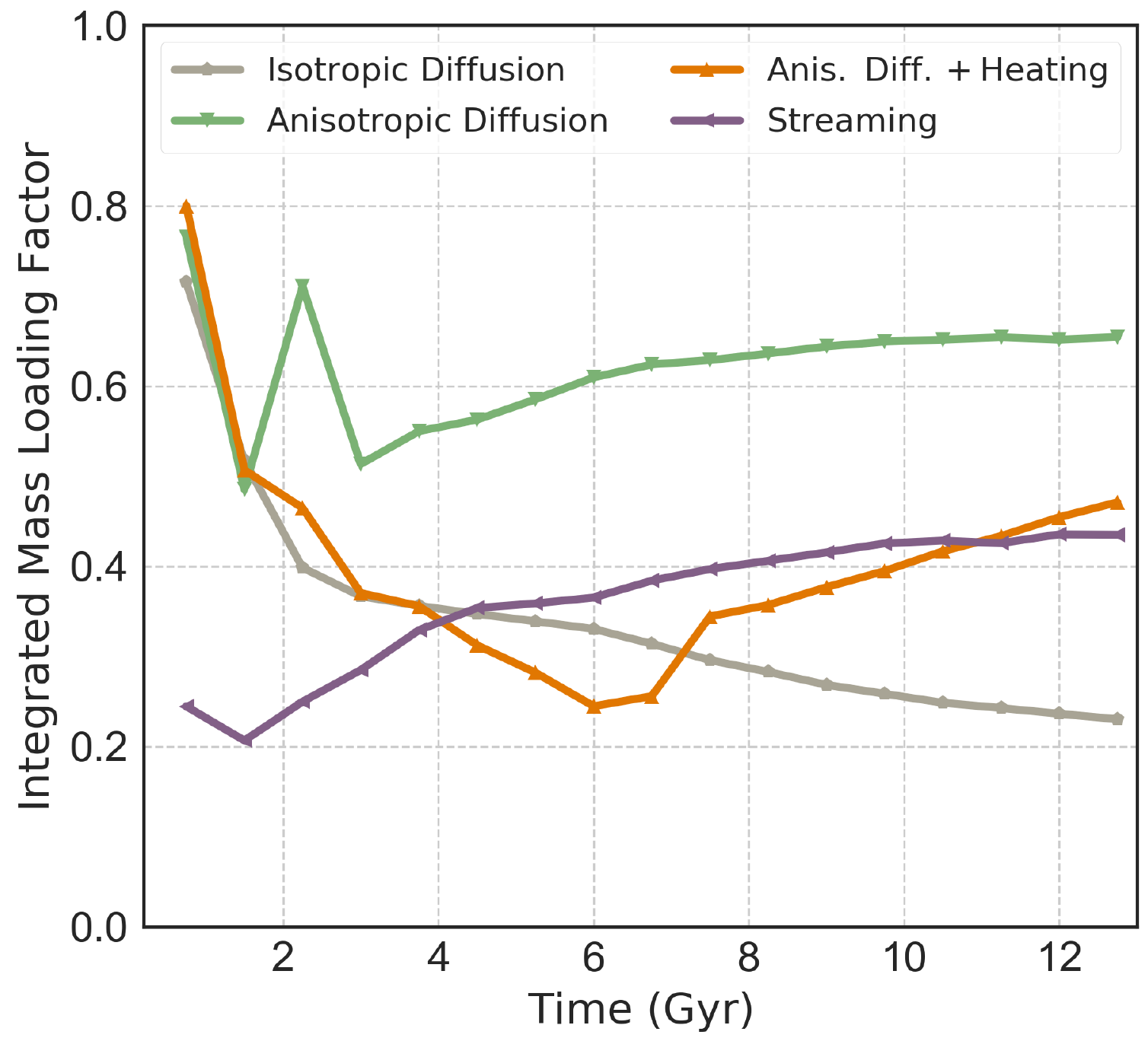}
\caption{\label{fig:massload}
The time-evolution of the integrated mass loading factor (the total ejected gas mass / total mass in stars), 
measured at 150 kpc from the galactic center.  
At late times, models with anisotropic CR transport continue driving gas out of the galaxy, even when star
formation is declining. } 
\end{center}
\end{figure}
\subsection{Initial Conditions}\label{sec:ICs}

We simulate a suite of idealized isolated disk galaxies with initial conditions
described by the AGORA collaboration \citep{Kim:2014}. The fixed dark matter halo
of mass $M_{200} = 1.074 \times 10^{12} M_{\odot}$ follows the 
Navarro-Frenk-White (NFW; \citealt{Navarro:1997}) profile and is situated in a
hot ($10^6$ K), stationary, uniform density box of (1.31 Mpc)$^3$.
The concentration and spin parameters are respectively defined to be 
$c = 10,$ $\lambda = 0.04$. 
The gaseous disk has a total mass of $M_d = 4.297\times 10^{10} M_{\odot}$
and follows the analytic exponential profile 

\begin{center}
$\rho(r,z) = \rho_0 e^{-r/r_d}e^{-|z|/z_d}$
\end{center}
where $r_d = 3.432$ kpc, $z_d = 343.2$ pc, and $\rho_0 = M_d f_{\mathrm{gas}}/
4\pi r^2_d z_d$. We define $f_{\mathrm{gas}} = 0.2$ to be the gas mass fraction 
of the disk. The rest of the disk mass (80\%) is contained in $10^5$ stellar 
particles. The stellar bulge has a stellar mass of $4.297 \times 10^9 M_{\odot}$
and follows the Hernquist density profile 
\citep{Hernquist:1990}.

We include an initial toroidal magnetic field of strength $\mathrm{B_0} = 1 \mu$G  
in the disk and a toroidal field of $\mathrm{B_0} = 10^{-15}$ G in the halo. 
The strong initial field in the disk follows the example of \cite{Ruszkowski:2017} to achieve
sufficiently fast CR streaming velocities at early simulation times. The strength of the initial halo
 field lies in the accepted theoretical range of primordial magnetic fields, 
$10^{-20} - 10^{-9}$ G \citep{Cheng:1994, Durrer:2013}.
To avoid numerical interpolation errors, galaxy models that
include CR physics are initialized with an isotropic 
background CR energy density of 0.1 erg/cm$^3$. 
This background CR pressure is
15 orders of magnitude weaker than the initial conditions of the gas pressure 
and does not alter the thermal gas in any significant way. 
Because we're interested in tracing the cycling process of metals, 
we set the initial metallicities of the disk and halo to be $0.3 
Z_{\odot}$ and $10^{-3} Z_{\odot}$ respectively.


\subsection{Description of Simulation Suite}\label{sec:suite}
Our simulation suite is designed to isolate the effect of different
implementations of CR transport. 
All of the galaxy models share the initial conditions described above. 
The fiducial models differ from each other only in their CR transport 
prescription and are described below: 
\begin{itemize}
\item Model {\sc ncr} does not include CR physics and serves as our control model. 

\item Model {\sc adv} only simulates CR advection with the bulk motion of the gas.

\item  Model {\sc isod} assumes isotropic CR diffusion using the algorithm described
in \citet{Salem:2014a} with a constant diffusion coefficient of
 $\kappa_{\varepsilon} = 3\times10^{28}$ cm$^2$ s$^{-1}$. 

\item Model {\sc anisd} assumes anisotropic CR diffusion along magnetic field lines
 with a constant diffusion coefficient of 
$\kappa_{\varepsilon} = 3\times10^{28}$ cm$^2$ s$^{-1}$. 

\item  Model {\sc anisdh} builds on model {\sc anisd} with an added 
CR heating term, $H_c$.  

\item  Model {\sc stream} assumes CR streaming with a streaming factor of $f_s = 4$
\citep{Ruszkowski:2017}.

\end{itemize}
All of our fiducial models that include CR physics (in bold font in Table \ref{tab:parameters})
inject 10\% of their supernova ejecta in the form of CR energy. 
See Table \ref{tab:parameters} for a summary of the different models.

\section{Results}\label{sec:results}
\begin{figure*}
\begin{center}
\includegraphics[width= \textwidth]{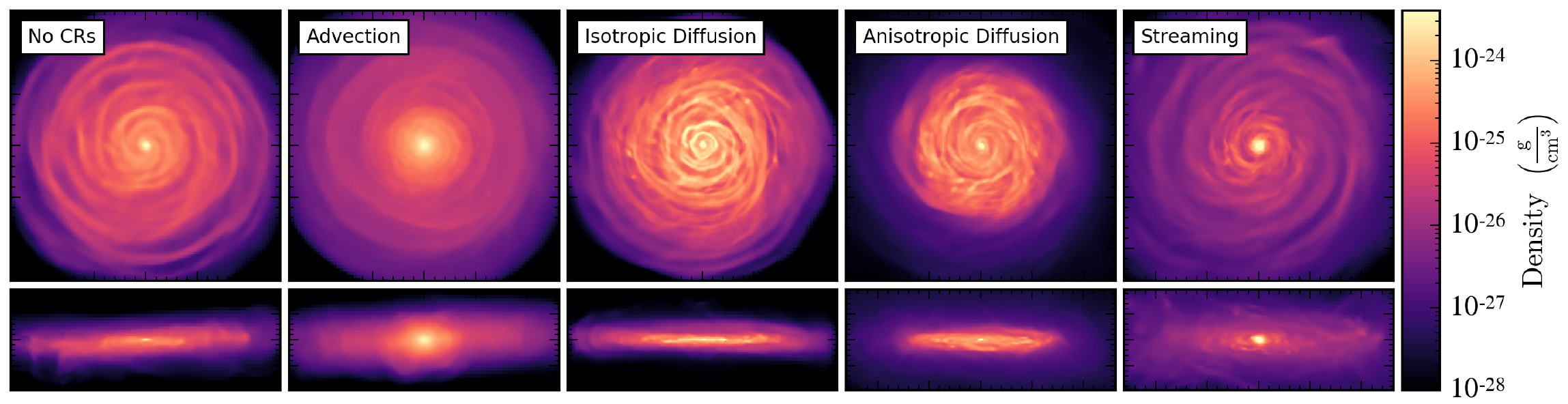}
\caption{\label{fig:density}
The face-on (top) and edge-on (bottom) density projections
of the gas in our fiducial models after 13 Gyr of evolution.  Although all 
galaxy models started from the same initial conditions, their morphologies 
depend on the CR transport mechanism. The control model, {\sc ncr}, is a MW-type
spiral galaxy.   Models {\sc adv} and {\sc stream} have extended gas profiles and 
thick disks.  Model {\sc anisd} has lost much of its gas to outflows. The top and bottom rows
show physical scales of 52 x 52 kpc and 20 x 52 kpc respectively.} 
\end{center}
\end{figure*}

\begin{figure}
\begin{center}
\includegraphics[width=0.45\textwidth]{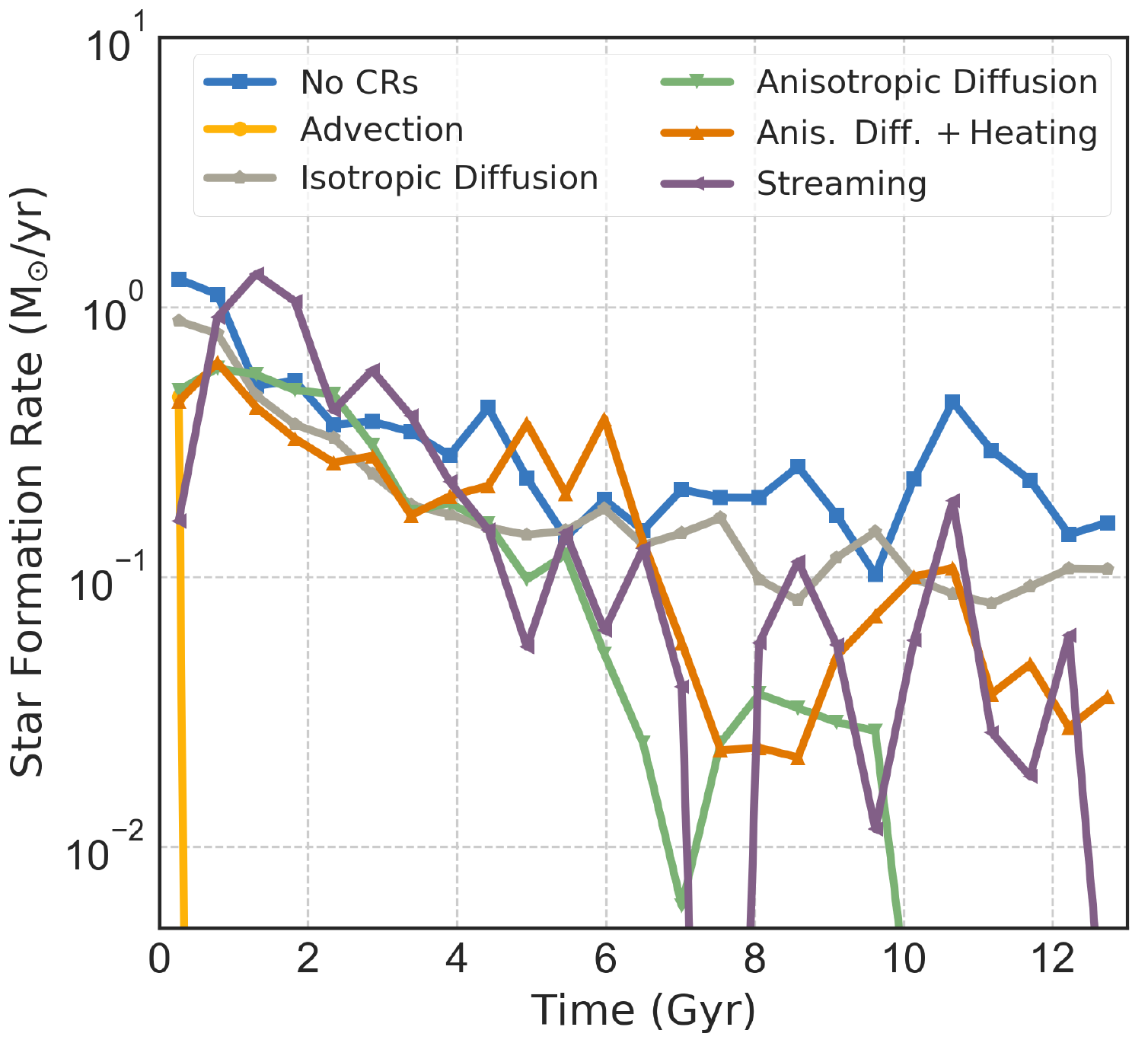}
\caption{\label{fig:sfr}
The star formation rate as a function of time for galaxy models with
different CR transport prescriptions. Model {\sc adv} quenches after the first episode of
star formation.  The star formation in model {\sc isod} consistently lies just below that of the
control model {\sc ncr} with no CRs. Model {\sc anisd} briefly matches the star formation in the
control around t = 2.5 Gyr, but is quenched soon after at around t = 10 Gyr. Model {\sc stream} has
a variable star formation history with a period of roughly 2 Gyr. }
\end{center}
\end{figure}

\subsection{Outflows}\label{sec:outflows}

The CGM in the isolated galaxy models is enriched solely by the outflows that
expel gas from the disk. For this reason, we first turn our attention to analyzing
the CR-driven winds and their dependence on the CR transport mechanism. 

In broad strokes, a CR-driven wind begins when CRs move down their gradient, 
out of the midplane of the galaxy. CR pressure support then lifts low-entropy 
gas out of the galactic potential well, triggering outflows. Figure \ref{fig:outflows} displays the relationship between 
gas entropy (top row), the vertical component of the velocity of the galactic winds
(middle row), and the ratio of CR to thermal pressures (bottom row) after 2 Gyr
of evolution. From left to right, the columns show models with:
No CR physics ({\sc ncr}), CR advection ({\sc adv}) only,  isotropic CR diffusion
({\sc isod}),  anisotropic CR diffusion ({\sc anisd}, and CR streaming ({\sc stream})).

Although models {\sc ncr} and {\sc adv} experienced a brief period of weak gas expulsion, by
t = 2 Gyr these galaxies have lost all signs of outflows and show no signs of strong inflows. 
This is reinforced by the gas entropy profiles, which resemble the initial conditions. The main difference
between models {\sc ncr} and {\sc adv} at this time is near the midplane of the galaxy, where
the CR pressure of model {\sc adv} has created a thicker vertical disk profile. 
Confined to move solely through advection, the CRs in model {\sc adv} have no 
efficient mechanism for escaping the disk. 

Models with CR transport relative to the gas ({\sc isod}, {\sc anisd}, and {\sc stream}) 
all drive strong outflows with velocities reaching $10^2$ km/s, consistent with 
previous works \citep[e.g.][]{Salem:2016,Pakmor:2016a,Wiener:2017}. 

Model {\sc isod} drives relatively thin and uniform conical outflows. 
The gas entropy profile is relatively unaltered, reaching values near $5\times10^2\ \mathrm{cm^2\ keV}$
near the midplane of the disk. CR pressure dominates over thermal pressure,
tracing the shape of the outflows out to a cylindrical radius of 10 kpc. Outside of the 
active outflow region, the CR pressure is below one tenth of the gas pressure.  

The outflows generated with anisotropic CR diffusion in model {\sc anisd} have a thinner
radial profile and reach higher velocities compared to those driven by isotropic diffusion. 
For $z > 20$ kpc outside of the midplane, there are signs of infalling gas surrounding the
outflow. The CR pressure dominates over thermal pressure for nearly all radii in the 
$100\times100$ kpc projection. The gas entropy reaches values of 50 cm$^2$ keV
just outside the disk, roughly an order of magnitude lower than that in model {\sc isod}. 

Compared to both of the diffusion models, the outflows generated by CR streaming 
start higher above the midplane of the disk and have a wider horizontal extent, reaching
radii of nearly 40 kpc at a vertical height of 50 kpc.  Near the galactic center, there are 
several filaments of inflowing gas. Compared to model {\sc anisd}, 
model {\sc stream} has lower gas entropy immediately outside the disk and at larger
radii, maintaining values around $5\times10^2$ cm$^2$ keV at radii 50 kpc.  Unlike either of the diffusion models, the distribution of the CR pressure
ratio in model {\sc stream} is patchy, ranging from 0.2 - 50 times that of the thermal pressure.

\begin{figure*}
\begin{center}
\includegraphics[width=\textwidth]{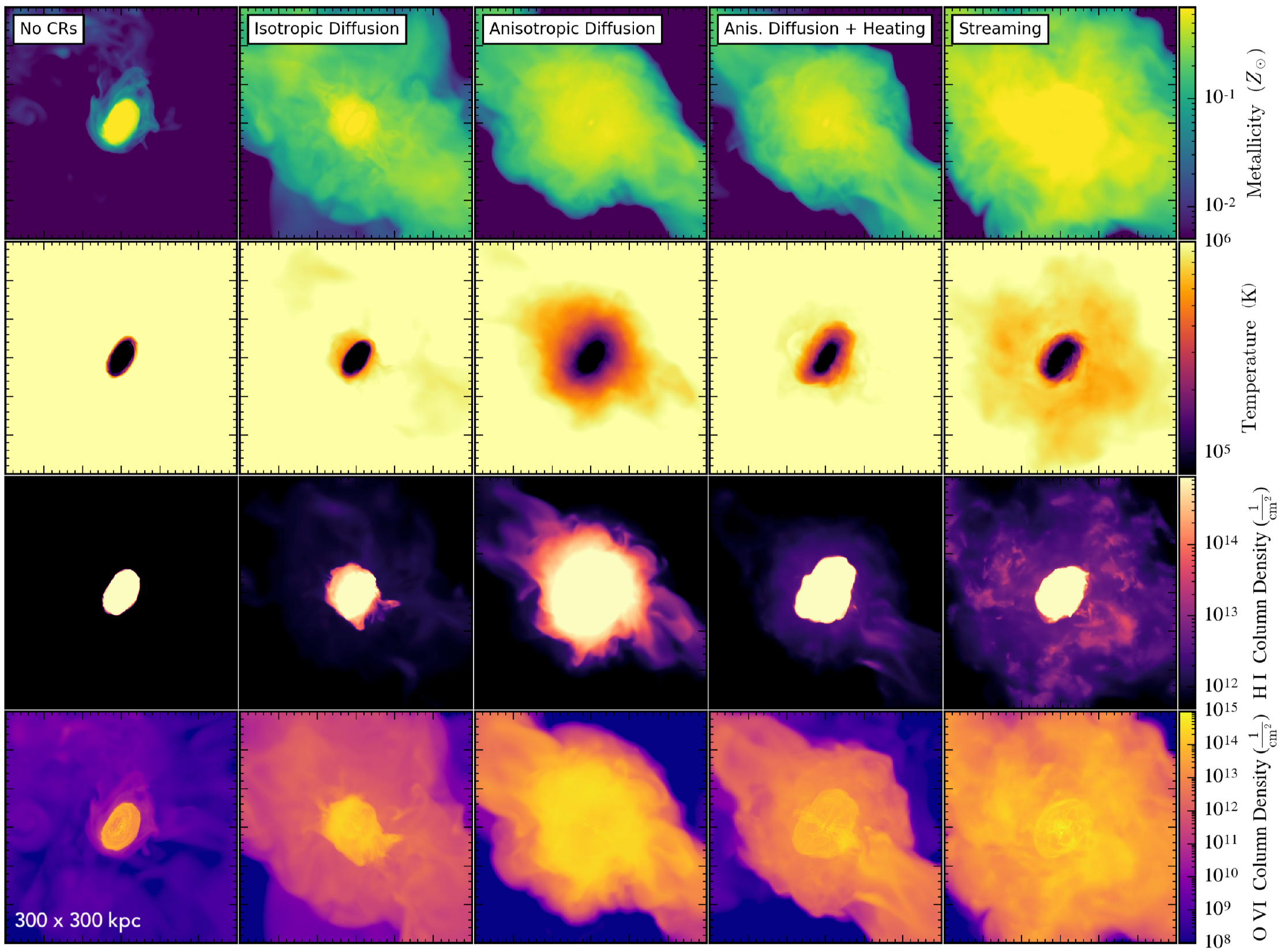}
\caption{\label{fig:ion_density} A comparison of the state of the CGM between the fiducial galaxy models
after 13 Gyr of evolution.
The top two rows show the density-weighted metallicity and temperature and
the bottom two rows show the unweighted column densities of H I and O VI.  Our control model {\sc ncr} drove
weak outflows, so the state of its CGM is relatively unaltered from the initial conditions. Model {\sc isod} has a  spatially uniform
distribution of warm, metal-rich, gas in its CGM. The galaxy model {\sc anisd } is surrounded by a reservoir of cool gas, traced
by H I. The temperature and ionization structure of model {\sc anisdh} follows the morphology of that in model {\sc anisd} although
the effect of CR pressure is weakened through losses due to CR heating.  Galaxy model {\sc stream} has a patchy, multiphase distribution
of different temperature gas, traced by its H I and O VI column densities. The projection is generated from a 300 kpc cube.}
\end{center}
\end{figure*}

Figure \ref{fig:velz} follows the time evolution of the density-weighted outflow velocity as a function of 
the height above the galactic disk, for models with different CR transport prescriptions. 
At early times, gas in our control model, {\sc ncr}, is inflowing at all radii within 200 kpc of the galactic
 center.  After 2 Gyr, weak ($\mathrm{v_z} = 10$ km/s) outflows develop at heights
 above 30 kpc. As we shall see in Figure \ref{fig:ion_density}, these weak outflows fail to enrich
 the CGM with metals. Instead, we turn our attention to models {\sc isod}, {\sc anisd}, and {\sc stream}, 
for which 10\% of supernova feedback was injected as CR energy. 

Model {\sc isod} has the strongest outflows at t = 1 Gyr, with radially-averaged velocities surpassing 
100 km/s. By t = 4 Gyr, these outflows weaken significantly, with average velocities hovering
 around 10 km/s. After 8 Gyr, notable inflows develop at radii above 30 kpc and
persist for the rest of the simulation time.  

This galactic wind is consistent with expectations of isotropic CR diffusion.
The initial burst of supernova feedback in model {\sc isod} creates a steep
 gradient of CR energy density. With isotropic diffusion, CRs drag gas out of the galaxy as they travel
 down their own gradient, uninhibited by the presence of magnetic fields. This process continues 
until the CR gradient flattens, slowing down the diffusion process. At later times, the star formation 
has decreased enough that the newly-injected CRs can no longer create a gradient steep enough to drive
 a strong wind. As the CRs in the CGM continue to stream away from the galaxy, they can't provide
 enough pressure support to the gas they expelled and it begins to collapse back down on to the 
galaxy. 

Model {\sc anisd} retains consistent outflow velocities within 50 kpc of the disk throughout 
its evolution. Beyond 50 kpc, the winds are sensitive to the star formation history.  For example, 
the double-valley feature at t = 8 Gyr traces the suppressed star formation at t = 7 Gyr
(see Figure \ref{fig:sfr}).

The steady nature of the galactic outflows in model {\sc anisd} is fueled by
anisotropic CR diffusion. In this approximation, the velocity with which CRs can escape the disk
is regulated by the complex geometry of magnetic field lines. Therefore, the galaxy releases
its built-up CR energy slowly over time. Although the galaxy is quenched after t = 10 Gyr, 
weak outflows in model {\sc anisd} persist out to t = 13 Gyr. 

Like in model {\sc anisd}, the key to sustained outflows in model {\sc stream} is its anisotropic
CR transport. Unlike any other fiducial run, model {\sc stream} shows signs of inflow
near the disk while simultaneously hosting outflows at larger radii. 

The integrated mass loading factor (IMLF), defined as the ratio of the total injected gas mass over the total stellar mass,
 quantifies the expelled gas content. 
Figure \ref{fig:massload} shows the evolution of the IMLF in 
models {\sc isod, anisd, anisdh,} and {\sc stream}. We exclude models {\sc nocr} and {\sc adv}, which 
didn't drive outflows, from this analysis. 

For the first 1.5 Gyr, the IMLF is nearly indistinguishable between models {\sc isod, anisd} and {\sc anisdh}. 
The low IMLF at early times in model {\sc stream} is most likely due to the later onset of the galactic wind as 
well as the wind's location above the galactic midplane. 

At later times, the IMLF decreases in model {\sc isod}, 
yet increases in the other three models. 
Model {\sc anisd} reaches the highest IMLF due to the accumulated
reservoir of CR energy near the galactic disk that continues to drive outflows
even after star formation has ceased. Models {\sc anisdh} and {\sc stream} have weaker 
IMLFs than model {\sc ansid} due to higher star formation rates and losses in CR energy through
the heating term.

\subsection{Star Formation and Morphology}\label{sec:morphology}
The relationship between CR transport and a galaxy's star formation rate (SFR) 
can dramatically influence a galaxy's morphology. 
The outflows generated by different CR transport prescriptions remove gas from the
disk which would have otherwise been available for star formation. 
In addition, the presence of strong CR pressure in the disk can prevent star formation
by stabilizing the thermal gas against collapse. In this section, we contrast the 
morphologies of our fiducial galaxy models after 
evolving the simulations for 13 Gyr. Because a galaxy's morphology is
so intricately tied to its star formation history, we describe Figures \ref{fig:density} and
\ref{fig:sfr} simultaneously.

Figure \ref{fig:density} displays the density-weighted face-on and edge-on projections of the gas density
in our fiducial galaxy models after 13 Gyr of evolution. In the left panel, model {\sc ncr} serves as the 
benchmark example of a MW-type disk galaxy. This galaxy has a clear spiral structure,
with the disk extending to a radius of roughly 25 kpc and a vertical height of roughly 3 kpc.  
The star formation rate of model {\sc ncr} begins with one solar mass per year and slowly decreases over time, 
hovering around a few tenths of a solar mass at later times.
With a modest star formation history and no galactic winds, 
model {\sc ncr} retains much of its gas in its disk, with average density values 
around $3\times10^{-25}$ g/cm$^3$.  

Although model {\sc adv} has had a similar outflow history as {\sc ncr}, its star formation history has
dramatically altered its morphology. Model {\sc adv} quenched only 200 Myr after its initial burst of star
formation (see Figure \ref{fig:sfr}).  Confined to advection-only transport, the CRs  in this galaxy
are trapped inside the galactic disk.  After the first episode of star formation induces CR feedback, 
CR pressure dominates over gas pressure (see the bottom-left panel of Figure \ref{fig:outflows} and 
the discussion in \S \ref{sec:outflows}). This additional pressure stabilizes the gas against 
against collapse, halting future star formation. 
Without galactic outflows to carry CRs out of the disk and without any new stars to create 
supernovae that could potentially drive outflows, this galaxy will remain quenched. 
Although model {\sc adv} has a substantial reservoir of gas in its disk, CR pressure
expands its vertical profile keeping the gas at lower average densities.

The star formation history of model {\sc isod} most closely follows that of model 
{\sc ncr}. Episodes of star formation that expel CRs into the disk keep the SFR of
model {\sc isod} consistently below that of the control. However, because isotropic
diffusion is efficient at removing CRs from the galactic disk, star formation is only  
weakly suppressed in this galaxy. 
The morphology of model {\sc isod} is qualitatively similar to that
of model {\sc ncr} out to a cylindrical radius of 15 kpc. Having expelled much of its 
ISM through outflows, model {\sc isod} has a shorter radial 
and vertical extent to its gas. 
\begin{figure*}
\begin{center}
\includegraphics[width=\textwidth]{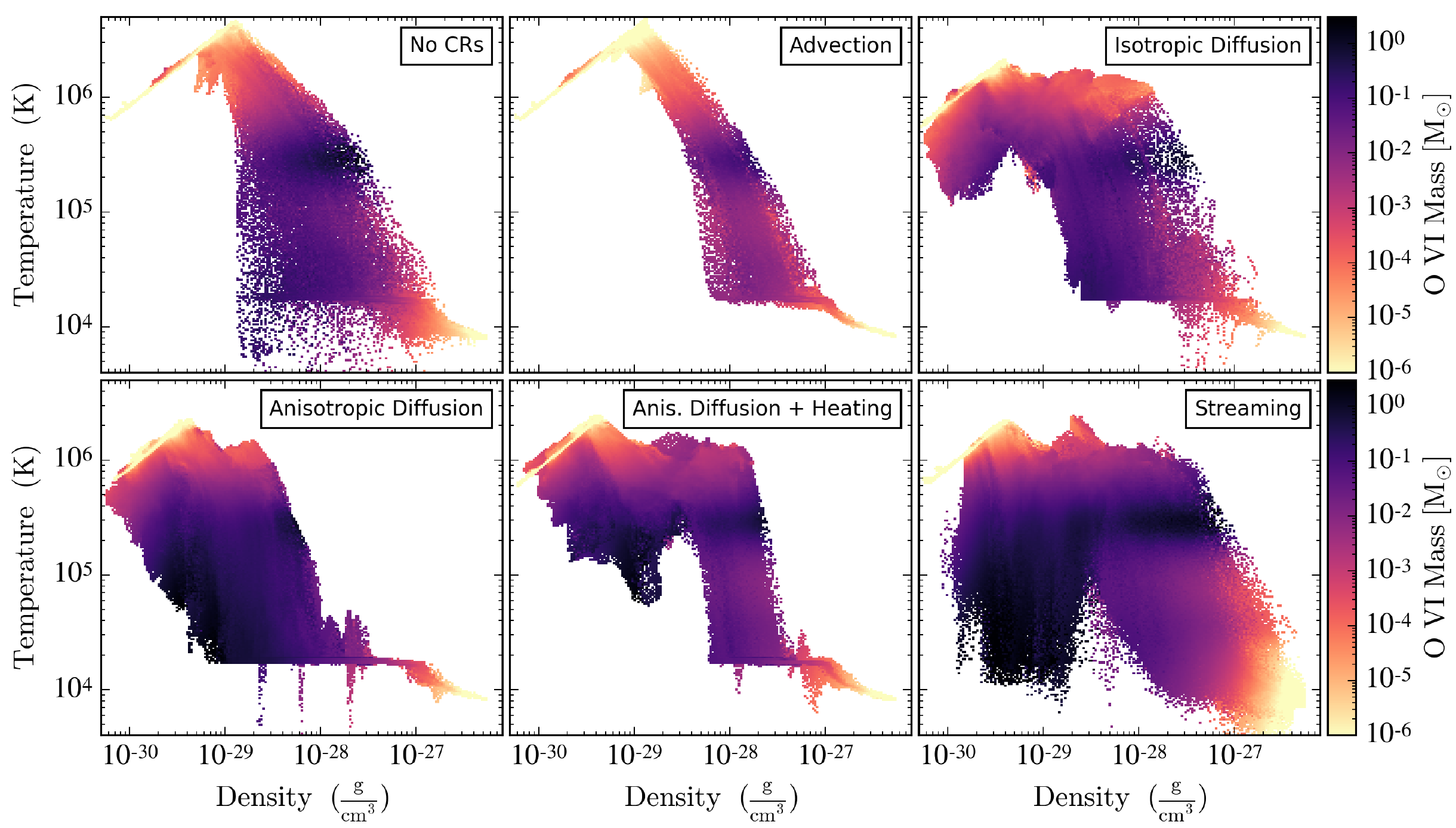}
\caption{\label{fig:dtm_phase}
The 2D histograms of temperature as a function of density for the six different
galaxy models after 13 Gyr of evolution. The color in the plot shows the mass of O VI in each bin. 
Model {\sc isod} has more diffuse gas at larger radii. 
Model {\sc stream} creates a more dramatic multiphase medium, with a larger range of possible 
densities for each temperature. Model {\sc stream} supports an abundance of O VI at both 
warm and cool temperatures. }
\end{center}
\end{figure*}

Anisotropic CR transport in model {\sc anisd} retains CR pressure in the disk, 
which suppresses star formation at early times. 
As galactic outflows develop, the CR pressure is lifted out of the disk, allowing for star
formation to resume.
After roughly 2 Gyr of evolution the star formation in model {\sc anisd} briefly matches
that of the control model. The injected CR pressure following this star formation
episode decreases future star formation, ultimately quenching the galaxy after 10 Gyr.
CR heating in model {\sc anisdh} relieves some of the CR pressure in the disk, leading
to a higher star formation rate than that in model {\sc anisd}.
Although at the time of the snapshot in Figure \ref{fig:density} model {\sc anisd} has been 
quenched for 3 Gyr, this galaxy model retains some spiral structure and high gas densities within
a 15 kpc radius of its center. The lingering CR pressure from past star formation creates an extended
low-density gas profile around the disk.

Both models {\sc anisd} and {\sc stream} have extended, thick gaseous profiles that are most apparent in the
edge-on view. Model {\sc anisd} has a bimodal distribution of gas, with a core around $2\times10^{25}$ g/cm$^3$
extending to 15 kpc, surrounded by a less dense gaseous halo extending to 25 kpc. Although some spiral 
structure is present, it is significantly less pronounced than in model {\sc isod}. Model {\sc stream} has relatively
low gas density in the disk with a spiral arm structure that is thinner than that of models {\sc ncr} or {\sc isod}.

Because of the toroidal geometry of the initial magnetic field, the CRs in model {\sc stream} suppress 
star formation for the first 500 Myr. Once the magnetic field develops a sufficiently strong vertical component,
 the CRs escape the disk, dragging thermal gas along with them. 
The outflows relieve the ISM of the CR pressure, allowing star formation to resume. 
Filaments of inflowing gas near the midplane of the disk (see  Figure \ref{fig:outflows}) supply
the additional gas necessary for extended episodes of star formation. 
Compared to the other galaxy models, model {\sc stream} has the lowest gas densities and the
most extended vertical and radial disk profile.


 \begin{figure*}
\begin{center}
\includegraphics[width=\textwidth]{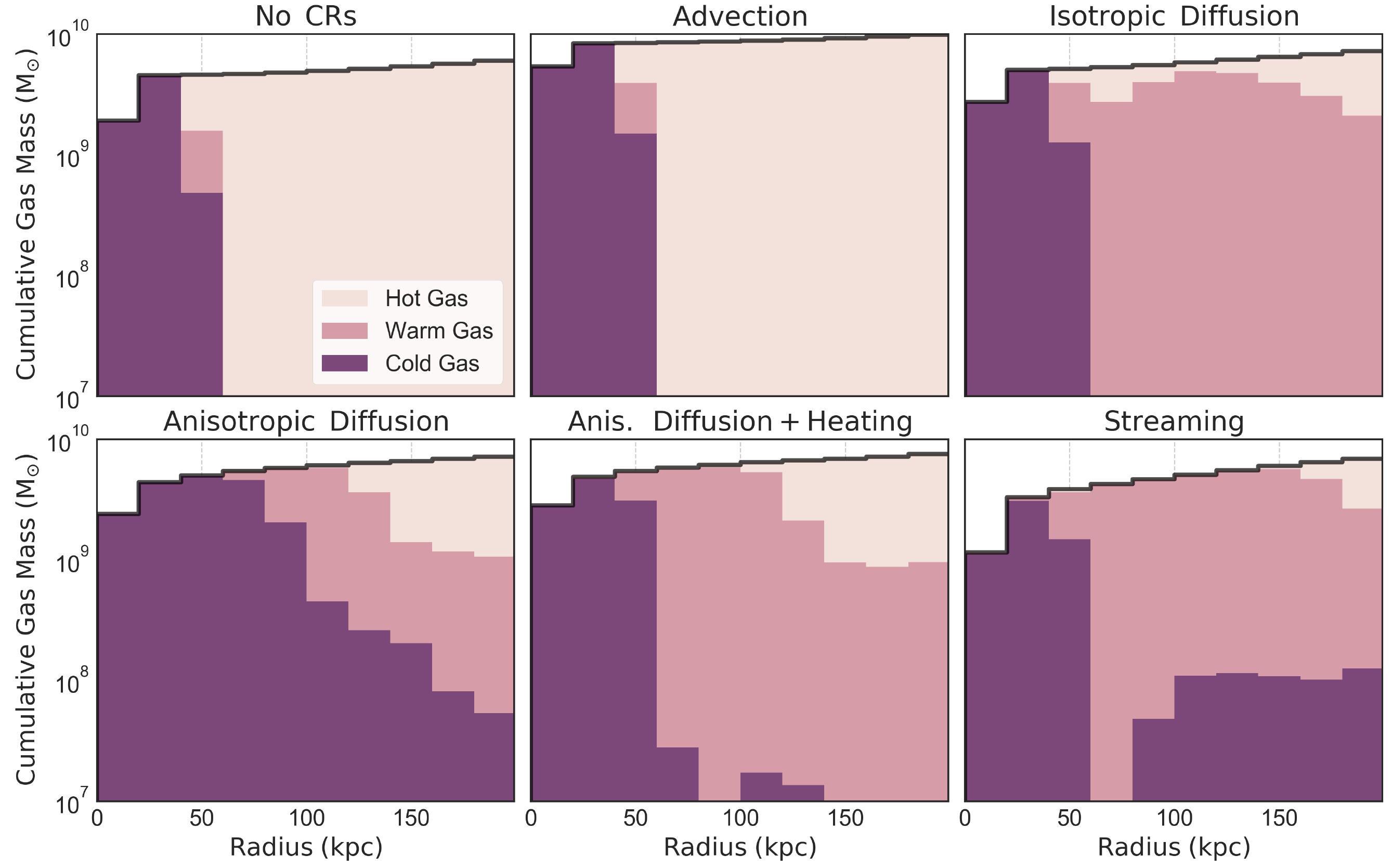}
\caption{\label{fig:temperature}
The black line follows total enclosed mass as a function of spherical radius for six different galaxy models. The colors of the bars 
denote the 
fractional distribution of the gas temperature at different radial shells. We consider three temperature bins of gas: cold gas ($T < 10^5 K$) in dark purple, 
warm gas ($10^5 K < T < 10^6 K$) in medium purple, and hot gas ($T > 10^6K$) in light purple. 
Only models with CR transport relative to the gas have a substantial amount of warm or cold gas in their CGM. 
Model {\sc anisd} has an abundance of cool gas near the galactic center that persists at large radii. 
Model {\sc stream} has a truly multiphase medium with an abundance of cold and warm at radii of 200 kpc. }

\end{center}
\end{figure*}

\subsection{The Simulated Circumgalactic Medium}

We now turn our attention to the different ionization structures within the CGM of the simulated
galaxy models.
Our control model, {\sc ncr}, has limited amounts of metals outside of the galactic disk. 
This is an unrealistic result and we will not dwell on its implications here. Our goal is to isolate the 
effects of CR transport on the CGM's temperature and ionization structure, and so the following 
figures and discussions will focus on models {\sc isod}, {\sc anisd}, and {\sc stream}.
Where relevant, we discuss model {\sc anisdh}, which includes the heating term $H_c$, that is 
present in the CR streaming prescription. Although the heating term is an unphysical addition to 
CR diffusion, its presence helps isolate which aspect of the CR streaming or diffusion mechanism
is responsible for observed differences in temperature and ionization states \citep{Wiener:2017}. 

Figure \ref{fig:ion_density} shows the density-weighted projections of gas metallicity
and temperature and the unweighted projections of H I and O VI column densities after 13 Gyr of evolution.
Each column holds a different mode of CR transport. Starting from the left, 
the columns show results for galaxy models {\sc ncr}, {\sc isod} , {\sc anisd}, {\sc anisdh}, and {\sc stream}. 

The outflows in models with CR diffusion or streaming have populated their CGM with metals
from the disk out to radii surpassing 200 kpc. Model {\sc isod} has a relatively uniform distribution of
metallicity in its CGM, with a density-weighted average value around 0.2 Z$_{\odot}$. 
Similarly, its temperature, H I and O VI column densities are also spatially uniform. 
CR pressure support creates a cooler temperature profile (roughly 
$3\times10^{5}$ K) compared to the control. 
The cooler temperatures result in  stronger column densities of H I and O VI compared to model {\sc ncr}. 

Anisotropic diffusion in model {\sc anisd} produces a steep temperature and H I column density gradient.
The temperature of the gas is coolest near the disk, where CR pressure is strongest, and decreases 
radially outwards. 
The H I column density follows the shape of the temperature profile, extending out to where the
density-weighted temperature is $10^6$ K. The O VI column density is sensitive to metallicity, 
so the column density profile traces the radial extent of the outflows. 
Compared to the other fiducial runs, model {\sc anisd} has the strongest column densities of H I and O VI. 
We point out that although this model has been quenched for 3 Gyr, the gradual release of CR pressure
results in the accumulation of cold gas just outside the disk.

The metallicity distribution of model {\sc anisdh} is qualitatively similar to that of model {\sc anisd}. 
However, the added heating term dramatically decreases the impact of CR pressure, resulting in
warmer temperatures and weaker column densities of H I and O VI than those in model {\sc anisd}. 

The outflows in model {\sc stream} reach larger radii than those in model {\sc anisd}. 
The temperature profile is dominated by clumps of  cool and 
warm gas extending out to radii of 150 kpc. 
Model {\sc stream} has a patchy distribution of both H I and O VI column densities, with both coexisting in
relative abundance at radii above 100 kpc from the galactic center. 

\begin{figure*}[t!]
\begin{center}
\includegraphics[width=\textwidth]{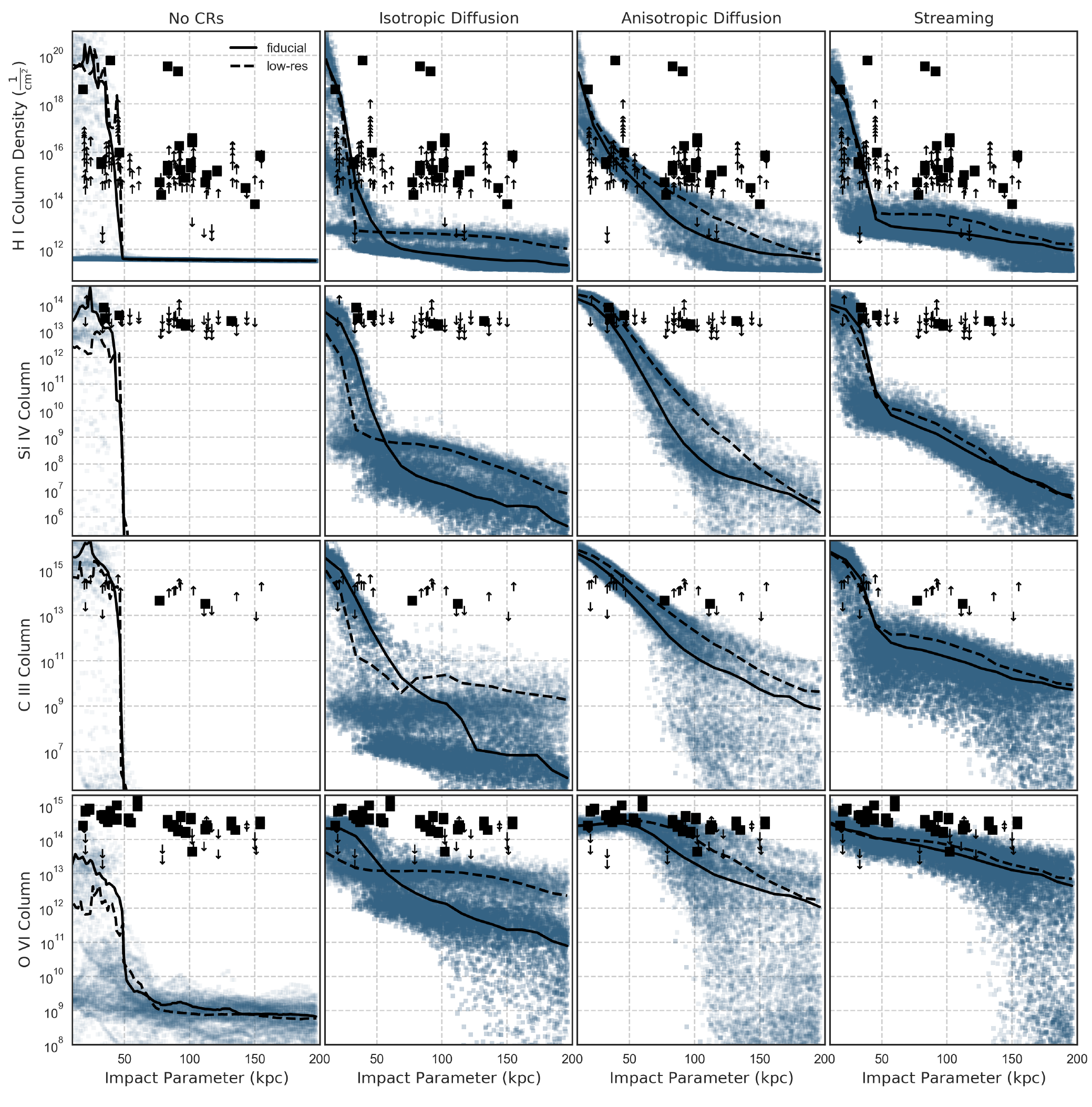}
\caption{\label{fig:col_density} Column densities of H I, Si IV, C III, and O VI as a function of impact parameter after 
13 Gyr of evolution. Each point on the graph shows a column density measurement using {\em Trident} from either
the fiducial run described by the title, or its low-resolution counterpart. The overplotted black lines show the average column
densities for the fiducial run (solid) and the low-resolution run containing the same physics (dashed). The black squares
and arrows represent observed column densities and upper and lower limits from \cite{Werk:2013}.
The control model underpredicts the observed column densities of all ions in this plot. Model {\sc isod} has a flat
 column density profile with respect to impact parameter. The column densities in models {\sc anisd} and {\sc anisdh}
 decrease with radius.  Model {\sc stream} predicts the strongest column densities and is least sensitive to changes in resolution. }
\end{center}
\end{figure*}
\subsection{Temperature Distribution}
To better understand the CGM structure, we 
analyze the abundance of O VI as a function of temperature and
density in Figure \ref{fig:dtm_phase}.
Since we are primarily interested with the gas in the CGM, we exclude data points
contained within a cylindrical radius of 25 kpc and a vertical height of 3 kpc of
the galactic center.
\begin{figure*}
\begin{center}
\includegraphics[width=\textwidth]{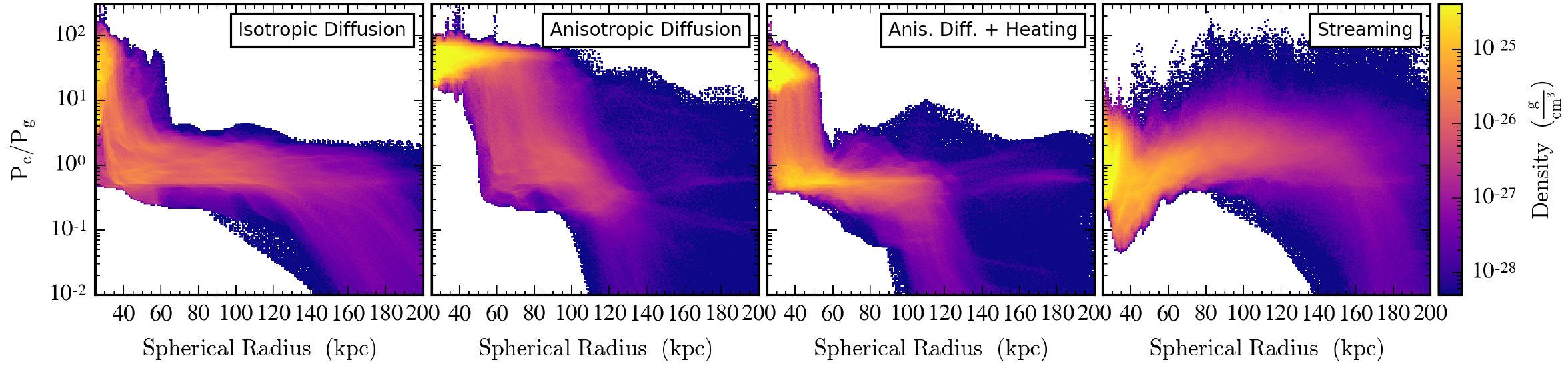}
\caption{\label{fig:phase_crbeta}
The unweighted ratio of CR pressure to gas pressure as a function of
spherical radius, measured from the galactic center after 13 Gyr of evolution. 
The color of the bins shows the density of the gas.
The disk has been removed from the dataset  in order
to focus on the CGM structure.  Model {\sc isod} has a relatively flat and unvarying profile of CR pressure 
in its CGM averaging around $\mathrm{P_{c}/P_{g}} \simeq 1$. Models {\sc anisd} and {\sc anisdh}  have
a wide range of CR pressure in the CGM, with  the highest values peaking near the disk center, at high gas 
densities. The CR pressure ratio in model {\sc stream} peaks at larger
radii, and at lower gas densities. }
\end{center}
\end{figure*}

The CGM of models {\sc ncr} and {\sc adv} has low masses of O VI and scarce amounts of low-density gas compared to 
the other models. Since low density values trace large radii in the CGM, the artificially high density floors 
are due to the lack of outflows. Suppressed star formation in model {\sc adv} is responsible for the low metallicities
in its disk that ultimately result in low masses of O VI.

Models with additional CR transport drive strong outflows that enrich the CGM with 
metals (see Figure \ref{fig:ion_density}). This enrichment creates higher column densities 
of O VI at large radii. However, the different
prescriptions for CR transport result in varied phase structures within the CGM. 

One such difference is the shape of the temperature-density phase diagram. 
Models with weak CR pressure support (models {\sc isod} and {\sc anisdh}) have
less cool gas at low densities than models {\sc anisd} and {\sc stream}. 
The strong CR pressure support in models {\sc anisd}
and {\sc stream} supports a wide range 
of temperatures at each density. This feature is
most pronounced at low densities in model {\sc stream}. 

CR transport also affects the abundance of O VI and the temperature
of the gas that produces it. 
Both models {\sc stream} and {\sc anisd} predict an
a reservoir of O VI ionized with the help of CR pressure support in low-density gas. 
However, only model {\sc stream} predicts an abundance of O VI photoionized with gas
temperatures around $3\times 10^5$ K, consistent with predictions in \citet{McQuinn:2017}.

Figure \ref{fig:temperature} shows the total enclosed mass as a function
of spherical radius for six different galaxy models. The colors of the bars 
denote the 
fractional distribution of the gas temperature at different radial shells. 
We consider three temperature bins of gas: cold gas ($T < 10^5 K$) in dark purple, 
warm gas ($10^5 K < T < 10^6 K$) in medium purple, and hot gas ($T > 10^6K$)
in light purple. The colored gas fraction is not cumulative
since the  cold mass contribution from the disk would dominate the distribution in 
some cases. 

Having driven weak outflows, model {\sc ncr} has trace amounts of cold or warm
gas outside of the disk. In contrast, although model {\sc adv} had similarly 
weak winds, the presence of CR pressure keeps the gas immediately around the 
disk below $10^6$ K. The influence of CR pressure drops abruptly after
a radius of 50 kpc. Because the CR pressure in the disk for this model 
stopped star formation almost immediately, model {\sc adv} has more gas 
available both inside and outside the disk of this model. 

Models in which CRs can move relative to the gas (via diffusion or streaming) 
have an altered temperature structure in their CGM. Model {\sc isod} is
dominated by warm gas out to large radii. Since isotropic diffusion depends 
solely on the direction and magnitude of the CR energy density gradient, 
the CR pressure distribution is nearly uniform throughout the halo. Over time, with
more stellar feedback releasing CRs into the disk, the CR pressure accumulates
in the halo, providing pressure support to the thermal gas. 

Models with anisotropic CR transport  exhibit a multiphase temperature 
structure. Models {\sc anisd} and {\sc anisdh} are both dominated by cold 
gas near the disk.  In both
models, the cold gas extends out to radii of 125 kpc and the warm
gas extends to radii of 200 kpc. 

Comparing this result with Figure
\ref{fig:ion_density}, we see the cool gas, traced by H I column densities, decrease
radially away from the center. The added heating term in model
{\sc anisdh} converts some CR pressure, which is responsible for supporting the 
warm and cold gas, into heating the gas, consistent with results in \cite{Wiener:2017}. 
Even with the heating term, the distribution of warm gas in model {\sc anisdh} is still
similar to that of {\sc anisd}. 

Model {\sc stream} shows signs of a true multiphase medium.
This model has the  lowest cumulative gas mass in its disk, 
likely due to its low gas densities and recent 
episodes of star formation (see Figure \ref{fig:sfr}). Although warm gas 
dominates its CGM out to radii of 150 kpc, cold gas survives in abundance
200 kpc away from the galactic center. 
Using Figure \ref{fig:ion_density}, we see that the temperature structure in
model {\sc stream} does indeed result in a patchy, multiphase medium.

\subsection{Ionization Structure}

Figure \ref{fig:col_density} holds the column densities of H I, Si IV, C III, and O VI as a function
of the spherical radius from the center of the galaxy. Each scatter point in the plot represents one
column density measurement at that radius. In each panel, we include points from both the fiducial
model and its low-resolution counterpart. The solid black lines show the average
column densities for the fiducial run while the dashed black lines show the average column densities
for the lower resolution run with the same physics. For details on how the column densities were 
constructed, see \S \ref{sec:trident}. 

To sufficiently sample our simulation space, we calculate the column densities at 
randomly-oriented sightlines passing through the CGM.
To generate a sightline, we first select a random point on the 
sphere of radius $r \in [10, 200]$ kpc from the galaxy center.  We then
chose a sightline by selecting a random angle in the plane tangent to the
sphere at that point. 

The control model, {\sc ncr}, underpredicts the column densities of all four ions. This is
both due to the lack of metals in its CGM and a deficit of cool gas. 

Of the remaining models, model {\sc isod} has the weakest column densities with
a flat distribution across impact parameter. This picture is consistent with 
weak and spatially uniform CR pressure in the CGM. 
Model {\sc anisd} has stronger column densities with a wider
spread at large radii. This is likely due to the uneven distribution of metallicity 
(see Figure \ref{fig:ion_density}). 
Model {\sc stream} predicts column densities that are similar in strength to 
model {\sc anisd}. However, the column density measurements in model {\sc stream}
are more tightly clustered around the average value, which is qualitatively similar to
observations. 

For all models, lower resolution predicts stronger column densities, possibly due to 
under-resolving the complicated structure of magnetic field lines. The column 
densities in model  {\sc stream} are the least sensitive to changes in resolution.

The column densities presented here are a qualitative example of the differences in ionization
structure between different models of CR transport. In reality, the exact values of ion column densities would
be influenced by the metallicity and inflows from the IGM. Therefore, in order to better match observations, we would
need to simulate galaxies in a cosmological context.


\begin{figure*}
\begin{center}
\includegraphics[width=\textwidth]{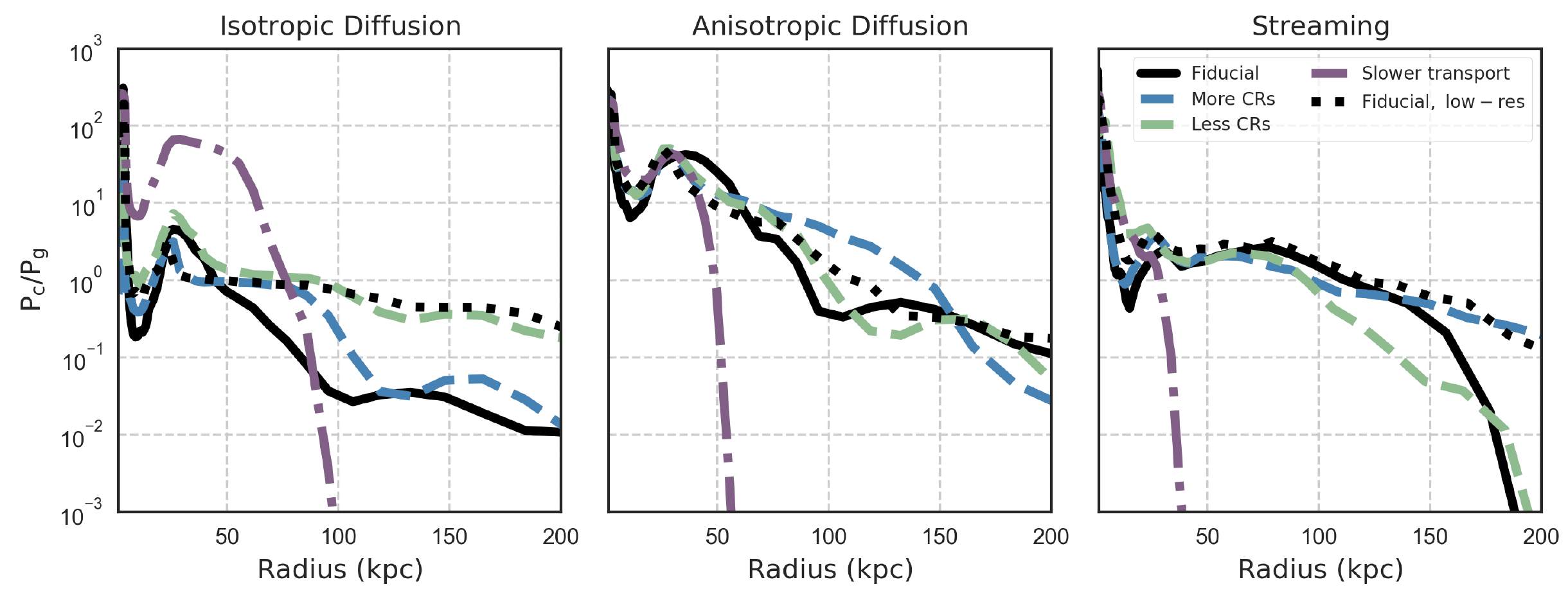}
\caption{\label{fig:CRBeta}
The density-weighted ratio of CR pressure to gas pressure as a function of
spherical radius after 3 Gyr of evolution. 
The black solid line represents the values for
the fiducial run of that CR transport method. Lines of different color and line style
show variations from the fiducial run as described in the legend. The blue and green 
lines represent galaxy models with CR injection fractions of 0.3 and 0.01 respectively.
The purple line represents a diffusion coefficient of $\kappa_{\varepsilon} = 10^{27} \mathrm{cm^2/s}$
for galaxy models {\sc isod} and {\sc anisd} and a streaming factor of $f_s = 1.0$ for 
model {\sc stream}.  The red line represents a diffusion coefficient of $\kappa_{\varepsilon} = 10^{29} \mathrm{cm^2/s}$
for model {\sc anisd}.  For a description of 
the simulation parameters see Table \ref{tab:parameters}. With this figure, we find that model {\sc isod} is the most 
sensitive to parameter changes while model {\sc stream} is the most robust. }
\end{center}
\end{figure*}
\subsection{The Distribution of Cosmic Rays in the CGM}
CR pressure in the CGM  impacts the temperature and ionization 
structure of the gas. For this reason, we investigate the role
of CR transport mechanisms in altering this CR pressure distribution. 

In Figure \ref{fig:phase_crbeta} we explore the distribution of CR pressure as a function
of spherical radius for models {\sc isod}, {\sc anisd}, {\sc anisdh} and {\sc stream}.
The pixels in the phase plot are colored by the density of the gas. Like in Figure
\ref{fig:dtm_phase}, we cut out the disk from our data sample.  

Model {\sc isod} has a flat and narrow profile of CR pressure as a function of radius, 
with the ratio $\mathrm{P_{c}/P_{g}}$ ranging between 0.5 and 5. 
At larger radii, the CR pressure ratio widens considerably, revealing regions where 
CR pressure is sparse. 

Model {\sc anisd} has a much wider range of CR pressure ratios across all radii. 
Compared to isotropic diffusion, anisotropic diffusion creates CR pressure ratios 
that are nearly an order of magnitude higher across all radii. In this model, the
CR pressure dominatesover thermal pressure near the galactic disk and at high gas densities.
The heating term in model {\sc anisdh} lowers the CR pressure ratio at all radii. 
However, this model still retains the  same qualitative shape of the CR pressure distribution
that is present in model {\sc anisd}.

Although model {\sc stream} also has a wide spread of CR to thermal gas pressure ratios, 
the distribution of that spread is qualitatively different from that in model {\sc anisd}. 
In this model, CR pressure in and near the disk is roughly 1.5 dex lower, spanning nearly 
2 dex in strength.  Unlike in model {\sc anisd}, the CR pressure is low near the galactic 
disk and dominates over thermal pressure at large radii (and low gas densities). 
\\
\subsection{Tuning the Knobs: The Impact of Choosing Parameters}
Figure \ref{fig:CRBeta} shows the density-averaged ratio of CR to thermal pressure as a function 
of impact parameter, measured after 3 Gyr of evolution. The solid black line
represents the low-resolution runs of our fiducial models.
All other lines differ from this run only in the parameter
specified in the label.  Blue and green lines represent galaxy models in which 
supernova feedback injected 1\% and 30\% of its energy as CRs, respectively. 
Purple lines represent slower CR transport velocities with either a diffusion 
coefficient of $\kappa_{\varepsilon} = 10^{27} \mathrm{cm}^2/s$
in models {\sc isod} and {\sc anisd} or a streaming factor of $f_s = 1.0$ in model 
{\sc stream}. To better compare models with anisotropic diffusion to those with isotropic
diffusion, we explore faster CR transport ($\kappa_{\varepsilon} = 10^{29} \mathrm{cm}^2/\mathrm{s}$)
in the middle panel, depicted by the red line. Finally, the black dotted line is the high-resolution fiducial run. 
For a summary of the parameters described above, 
see Table 1. 

Varying the injected CR fraction does not correspond with a linear change in the ratio of 
CR to thermal pressure in the CGM. For example, increasing $f_{CR}$ to 0.3
only marginally increases CR pressure in model {\sc isod}.  Counter-intuitively, increasing
$f_{CR}$ in models {\sc anisd} and {\sc stream} actually decreases the
CR pressure ratio in the CGM by suppressing star formation in the disk and thus
preventing the injection of additional CRs.  On the 
flip side, decreasing
the injected CR fraction by a factor of 10 from the fiducial value does decrease the
CR pressure in model {\sc isod} by roughly an order of magnitude at radii above 100 kpc. 
However, this decrease in $f_{CR}$ does not significantly 
alter the CR pressure ratio for models {\sc anisd} and {\sc stream}. 

The CR pressure distribution in all three models depends strongly on 
their transport velocities. Decreasing
the diffusion coefficient by a factor of 10 limits the reach of outflows to
100 kpc in model {\sc isod} and 60 kpc in model {\sc anisd}. Similarly, 
decreasing the streaming factor by four limits the reach of CR pressure
to 40 kpc in 3 Gyr. In addition to limiting the radial extent of CR pressure, 
the decrease in transport velocity in model {\sc isod} increases the CR pressure
ratio by a factor of 10 at small radii.  Increasing the transport velocity in
model {\sc anisd}  decreases the CR pressure within 75 kpc
of the galactic center, but has no discernible effects at larger radii.

Doubling the resolution decreases the CR pressure ratio by roughly
an order of magnitude for models {\sc isod} and {\sc anisd}. In model
{\sc stream}, the CR pressure ratio of the higher resolution run only deviates
from the fiducial run above 150 kpc. 

Model {\sc isod} is the most sensitive 
to changing parameters. In this approximation of CR transport, CRs can only 
move relative to the gas by diffusing down their own gradient. This motion is very 
sensitive to the injection fraction of CRs, which defines the strength 
of the CR gradient. Anisotropic diffusion is moderated by transport along magnetic
field lines. The complicated geometry of the magnetic field lines slows anisotropic 
diffusion near the disk, making model {\sc anisd} less sensitive to the CR injection fraction 
and star formation history than model {\sc isod}. 
The most robust model is {\sc stream}, in which the CR transport depends on
the CR gradient, magnetic field strength, and Alfv{\`e}n wave velocity. 
The CR pressure ratio in all three models decreases with slower CR transport and 
increased resolution. Breaking this degeneracy will require additional parameter studies.


\section{Discussion}\label{sec:discussion}

With our suite of isolated disk galaxy simulations, we have shown that 
the density, temperature, and ionization structures of the simulated CGM depend 
on the choice of CR transport mechanism. This discrepancy stems from the nature
of the CR transport approximation and cannot be trivially remedied by altering constant 
parameters. In the following section, we will summarize the qualitative nature
of each CR transport prescription and discuss potential improvements for
future simulations.  

The mere presence of CRs alongside thermal gas is not enough to drive
galactic outflows. This point is demonstrated with model {\sc adv}, in which
CRs are confined to propagate only through advection with the bulk motion 
of the thermal gas.  
These CRs provide pressure support to the gas within the disk but have no
efficient mechanism through which to escape into the CGM. This added 
pressure stabilizes the gas against collapsing to form stars and thus
quenches the galaxy almost immediately after the first episode of star formation. 
Although some CRs ultimately do escape the disk, there is no significant
CR influence in the CGM after 13 Gyr of evolution.  

CR transport {\it relative} to the thermal
gas is necessary to drive galactic winds or to reproduce the observed
column densities of ions in the CGM.  As CRs move out of the disk, 
they provide pressure support which lifts low-entropy gas out of
the gravitational potential well (see Figure \ref{fig:outflows}). 
This additional CR pressure lowers the temperature and increases
the ionization state of low ions, such as C III and Si IV, at large radii. However, the CGM
structure depends on the distribution of CR pressure, which 
varies substantially between different CR transport models.

In the isotropic diffusion approximation, CRs move down their energy gradient
with a velocity that is determined by both the strength of the gradient and the 
constant diffusion coefficient, $\kappa_{\varepsilon}$. 
The resulting galactic outflows and CGM are therefore sensitive to recent
star formation, which injects CR energy into the ISM.
The newly-ejected CRs propagate away from the galactic disk
 until the CR energy gradient is sufficiently flattened. 
Therefore, at late times,  additional CR pressure injected by 
ongoing star formation can no longer drive strong outflows. 
Ultimately, the CR pressure in the CGM is unable to support the
gas that it expelled at earlier times, triggering inflows. 
Varying the value of the constant diffusion coefficient alters the time scales on 
which the CR energy gradient flattens and the radial extent of CR energy, but 
not the qualitative shape of the CR distribution. 

Although isotropic diffusion is a crude approximation to the true interactions
between CRs and magnetic fields, it is a computationally frugal choice as it circumvents
the need for fully MHD galaxy models. Simulations with isotropic CR diffusion have 
been successful at driving strong outflows and increasing the column densities of
low ions in the CGM \citep{Salem:2014a, Wiener:2017}.
However, in our simulations, model {\sc isod} was the least effective at producing a 
multiphase medium and the most sensitive to the choice of constant parameter values.  

Anisotropic diffusion improves upon isotropic diffusion by approximating CR transport
as a random walk down the CR energy density gradient, along the magnetic field. 
This transport around magnetic field lines modulates the velocity of CR transport. 
Near the disk, where the magnetic field lines
are the most tangled, some CRs become trapped in their motion around magnetic field loops
and make slow radial progress away from the disk. Simultaneously, CRs near
magnetic field lines pointing out of the disk escape to larger radii. 
Therefore, anisotropic diffusion is capable of driving large-scale outflows 
while keeping a substantial presence of CR pressure near the disk. 
In this model, CR pressures are strongest near the galactic center and 
decrease rapidly at large radii. 
This CR pressure suppresses star formation in the disk and supports an abundance
of cool gas in the the halo. The strength and radial extent of the CR pressure depends
on the constant diffusion coefficient, magnetic field topology, and the star formation history.

Increasing the diffusion coefficient in models with anisotropic diffusion  
counteracts the effect of tangled magnetic field lines near the galactic center. 
However, the resulting models are still not directly comparable to models with 
isotropic diffusion with a lower diffusion coefficient.  
The difference lies in the variable timing of the CR propagation. In anisotropic diffusion models, the CR
pressure distributes itself preferentially along magnetic field lines, so the nominally constant diffusion
coefficient becomes a function of the magnetic field geometry. 
There is no single ratio of diffusion coefficients that results in the same distribution of
CR energy in both anisotropic and isotropic diffusion models.

CR streaming is the first-order approximation to CR transport.
Streaming CRs move along magnetic field lines with a velocity that depends
on the shape, {\it direction}, and {\it strength} of the magnetic field.  
This mode of transport creates a CR distribution in the CGM that supports  a  
truly multiphase medium, with cold gas clumps surviving alongside a 
warm and hot medium. 
We find that compared to diffusion, streaming is less sensitive to changes 
in star formation history or the choice of constant parameter values. 

A key component of the streaming approximation is the additional heating term, 
through which CRs give up their momentum to heat thermal gas.  
This transfer prevents simulations from overestimating CR pressure in the disk and halo. 
To discern the effect of the heating term from the streaming behavior, we
simulated a galaxy with both anisotropic CR diffusion and CR heating \citep{Wiener:2017}.
Although the additional heating term increased temperatures and lowered the column densities
of H I and O VI, the CR distribution in the CGM remained qualitatively the same.  
Therefore, we conclude that the key differences between the streaming and diffusion 
models lie in the transport approximations.

\subsection{Limitations and Future Work}

In this work, we explored the qualitative impact of the choice of CR transport on 
the simulated CGM. However, before simulations with CR physics can hold 
predictive power, several improvements must be made to constrain the 
details of this transport.  We discuss these factors in greater detail below. 

\subsubsection{Magnetic Fields}
Magnetic fields are the media along which CRs propagate and are therefore  crucially 
important for robust implementations of CR physics. The shape of the magnetic field 
dictates CR transport in the anisotropic diffusion and streaming approximations
while the magnetic field strength sets the streaming velocity and the rate of heat
transfer from CRs to the thermal gas.  In addition, magnetic pressure is inversely proportional
to the length scale of thermal instabilities in the CGM \citep{Ji:2017}. 
Therefore, simulating CR transport and recreating the multiphase CGM requires 
a realistic treatment of magnetic fields.  

Simulating the co-evolution of CRs and magnetic fields is complicated at early 
simulation times. The CRs streaming velocity depends on the magnetic field strength, 
yet primordial fields are believed to be no stronger than $10^{-9}$ G  \citep{Cheng:1994, Durrer:2013}.
If newly-injected CRs cannot escape the galactic disk before the next round of star formation, the galaxy 
risks becoming immediately quenched. 

One way to remedy this is to include magnetic supernova feedback, which can fuel the 
exponential magnetic field growth to observed values \citep{Butsky:2017}. 
However, in our attempts to simulate such a field, the small supernova injection sites were not sufficiently
resolved to accurately capture CR transport. The overabundance of CR energy in the disk
suppressed star formation, which in turn suppressed additional magnetic field injection. 
In these test simulations, the magnetic fields never reached observable strengths, 
and the streaming CRs never escaped the disk.

To circumvent this problem, we set the initial disk magnetic field to  $B_{0,d} = 1 \mu$G, 
similar to that in \citet{Ruszkowski:2017}. This initial field is strong enough 
for CR streaming models to drive outflows that are comparable in their reach to outflows 
generated by the diffusion models with $\kappa_{\varepsilon} = 10^{28}\mathrm{cm^2/s}$. 
The initial magnetic field in the CGM, which is the focus of this work, was set to $B_{0,h}=10^{-15}$ G. 
Therefore, the evolution of the of the magnetic field in the halo was driven by magnetized
galactic winds and the turbulence they produced.

Improved implementation of the co-evolution of magnetic fields with streaming CRs 
will require detailed resolution and parameter studies which are outside of the scope of this work.

\subsubsection{Improved CR Physics}
Our work makes several assumptions that are commonplace in 
current implementations of CR transport. 
One such approximation is that of a constant diffusion coefficient,
$\kappa_{\varepsilon}$, which avoids calculating the 
momentum-weighted integral of the local CR energy density. 
The velocity of CR transport in the diffusion limit is therefore very sensitive to 
the choice of $\kappa_{\varepsilon}$ (see Figure \ref{fig:CRBeta}).
Because the value of $\kappa_{\varepsilon}$ is poorly constrained, simulations
with the same model for CR transport can produce significantly varied results.

However, it is possible to improve upon the constant diffusion model
without the computational expense of explicitly solving for the
momentum-weighted value at every grid cell.
For example, \citet{Farber:2017} recently demonstrated that 
using a temperature-dependent, bimodal value for $\kappa_{\varepsilon}$ 
alters galactic wind properties and spatial distribution of CRs.

The underlying assumption of our CR transport models is that
CRs must propagate either in the diffusion limit or the streaming limit. 
Realistically, both modes of transport must be taking place. 
In a recent advancement, \citet{Jiang:2018} describe a new numerical scheme for CR
transport that self-consistently handles both streaming and diffusion processes. 
This prescription is a promising new direction that may resolve the discrepancies between
current implementations of CR transport presented in this work.

\subsubsection{Cosmological Context}
The idealized experimental setup of isolated disk galaxies 
offers great control in isolating the effects of CR-driven galactic
outflows on the structure of the CGM. However, this isolation
comes at the expense of simulating a realistic CGM.

For example, the CGM in our simulations is populated almost entirely 
by CR-driven outflows.  Neglecting inflows from the IGM leads to
an unrealistically empty CGM, which is particularly apparent in our control model. 
The isolated disk galaxy model is an oversimplification of galactic evolution since most galaxies, 
including our own Milky Way, evolve in a group or cluster environment. 
Interactions with nearby galaxies and satellites can have a significant
effect on the host galaxy's gas supply, which in turn affects its ability
to form stars and drive outflows. These interactions can also strip 
gas directly from the CGM.

We focus solely on the CGM of Milky Way-type galaxies, in 
which purely thermal feedback is successful at reproducing
galactic disk properties.
However, the differences between CR transport mechanisms
presented above are may change with galaxy mass.
For example, \citet{Jacob:2017} recently showed that the strength and mass-loading factor galactic outflows
driven by isotropic CR diffusion depends on the host galaxy's mass. 
This mass-dependency is likely to be present for anisotropic diffusion and streaming transport 
prescriptions. However, the nature of that relationship may change for different transport mechanisms. 

Future parameter studies using cosmological simulations will be necessary to  
develop a robust prescription of CR transport. We judge our CR transport models
by the temperature and ionization structure of the simulated CGM. However, CR
collisions in the CGM are expected to account for up to 10\% of the diffuse, isotropic gamma ray background
 \citep{Feldmann:2013}.  Therefore, observations of the diffuse gamma ray emissions, such as those 
taken with the Fermi-LAT telescope can be used to constrain CR transport models
\citep{Ackermann:2012}.

\section{Summary}\label{sec:summary}

Simulations including CR feedback are more effective at driving galactic
outflows and reproducing the observed ionization structure in the CGM 
than models with purely thermal feedback. However, models for
simulating CR feedback and transport are poorly constrained.

In this work, we demonstrate that galactic outflows and CGM structure  
are sensitive to the invoked CR transport mechanism. 
We achieve this by simulating a suite of isolated MW-type 
disk galaxies with three commonly-used prescriptions for CR transport
relative to thermal gas:
isotropic diffusion, anisotropic diffusion, and streaming. For completeness, 
we also include advection-only models, in which CRs are constrained to move with 
the bulk motion of the gas, and control models without any CR physics.
The results are summarized below. 

\begin{itemize}
\item  Models with CR transport relative to the thermal gas (streaming or diffusion) drive strong outflows.  These outflows
are generated by CR pressure support, which lifts thermal gas higher out of the gravitational potential well. 
Models with no CR feedback and those with CR advection do not launch strong
galactic winds.  

\item Models with isotropic diffusion launch the fastest winds that last around 5 Gyr.
 Models with anisotropic CR diffusion and streaming drive steady winds that persist after 13 Gyr 
of evolution. Models with anisotropic diffusion continue to drive outflows even after the 
galaxy is quenched. Models with CR streaming support simultaneous inflows near the disk
and outflows at larger radii. 

\item All models with CR feedback suppress star formation by supporting thermal
gas against collapse.  The degree
to which star formation is suppressed depends on the CR pressure within the disk.    
Models with pure CR advection quenched after the first episode of star formation. 
The star formation in models with isotropic diffusion closely followed that of the control. 
Models with anisotropic diffusion suppressed star formation more efficiently, 
ultimately quenching after 10 Gyr. Models with CR streaming had
a cyclical star formation history, supported by inflows from the CGM.

\item CR pressure in the CGM supports gas with cooler temperatures. However, that temperature
structure is sensitive to the CR transport model. 
The CGM of models with isotropic diffusion is primarily composed of warm gas
that is spatially uniform.
Models with anisotropic diffusion produce large quantities of cool ($T < 10^{5} K$)
gas out to radii of 100 kpc that remain even after star formation is quenched. This is an
interesting example of a quenched galaxy with a reservoir of cool gas in its CGM 
(e.g., \citet{Gauthier:2010,Thom:2012}). Models with CR streaming produce a patchy, 
multiphase gas distribution with cool gas existing alongside warm and hot gas
at large radii. 

\item CR pressure creates a multiphase medium, allowing gas of 
a given temperature to span several orders of magnitude
in density. At late times, models with isotropic diffusion
show a relatively uniform CGM temperature, whereas anisotropic diffusion and 
streaming models retain varied temperature profiles. In anisotropic diffusion models,  
the influence of CRs is strongest near the galactic center and decreases with spherical
radius. In CR streaming models, the multiphase temperature structure is patchy, with
clumps of cool gas existing 200 kpc from the galactic center.  

\item Models with anisotropic diffusion and streaming generate higher column 
densities of H I, Si IV, C III, and O VI than models with isotropic diffusion. 
The column densities generated by models with CR streaming have less variation in predicted
column densities as a function of impact parameter and are less sensitive to changes in resolution.

\item The differences between our galaxy models stem from the varied distribution
of CRs in the CGM. Compared to isotropic diffusion, anisotropic diffusion and CR streaming are more effective at 
retaining CR pressure near the galactic disk.  Since the transport in isotropic diffusion depends 
only on the gradient of CR energy density, its CR pressure support in the CGM decreases at late simulation times.

\item The distribution of CR pressure in the CGM is sensitive to runtime parameters such as
the amount of CRs injected in supernova feedback, the velocity of CR transport, and the resolution. 
We find that models with isotropic CR diffusion are the most sensitive to changes in these parameters.
This is because the velocity of CR transport in this approximation depend only on the CR energy 
gradient and the choice of constant diffusion coefficient.  The CR pressure distribution in models 
with anisotropic diffusion are less sensitive in comparison. 
We conclude that models with streaming, in which CR transport depends on the shape, direction, and strength
of the magnetic field lines, are the most robust. 

\end{itemize}

We have demonstrated that CR feedback can drive strong galactic outflows and 
provide the necessary pressure support to reproduce the multiphase temperature and ionization
structure of the CGM. 
However, because the state of the simulated CGM depends strongly on the invoked CR transport 
method, it is necessary to first develop a robust numerical CR transport model before simulations 
with CR feedback can hold predictive power.  

\acknowledgments
\section{Acknowledgments}
The authors would like to thank the anonymous referee for their insightful
suggestions.
They also thank Cameron Hummels, Julianne Dalcanton, Juliette Becker, 
Jessica Werk,  Matt McQuinn, 
\u{Z}eljko Ivezi\'{c}, Victoria Meadows, and Scott Anderson
for their valuable comments on this manuscript.  
I.B. also thanks Daniel Sotolongo for many
helpful conversations. I.B. was supported by the National Science
Foundation (NSF) Blue Waters Graduate Fellowship. This
research is part of the Blue Waters sustained-petascale
computing project, which is supported by the National Science
Foundation (Grants No. OCI-0725070 and No. ACI-1238993)
and the State of Illinois. Blue Waters is a joint effort of the
University of Illinois at Urbana-Champaign and its National
Center for Supercomputing Applications. 
\bibliographystyle{apj}
\bibliography{ButskyReferences}

\begin{thebibliography}{}
\expandafter\ifx\csname natexlab\endcsname\relax\def\natexlab#1{#1}\fi

\bibitem[{{Abadi} {et~al.}(2003){Abadi}, {Navarro}, {Steinmetz}, \&
  {Eke}}]{Abadi:2003}
{Abadi}, M.~G., {Navarro}, J.~F., {Steinmetz}, M., \& {Eke}, V.~R. 2003, \apj,
  597, 21

\bibitem[{{Ackermann} {et~al.}(2012){Ackermann}, {Ajello}, {Atwood}, {Baldini},
  {Ballet}, {Barbiellini}, {Bastieri}, {Bechtol}, {Bellazzini}, {Berenji},
  {Blandford}, {Bloom}, {Bonamente}, {Borgland}, {Brandt}, {Bregeon},
  {Brigida}, {Bruel}, {Buehler}, {Buson}, {Caliandro}, {Cameron}, {Caraveo},
  {Cavazzuti}, {Cecchi}, {Charles}, {Chekhtman}, {Chiang}, {Ciprini}, {Claus},
  {Cohen-Tanugi}, {Conrad}, {Cutini}, {de Angelis}, {de Palma}, {Dermer},
  {Digel}, {Silva}, {Drell}, {Drlica-Wagner}, {Falletti}, {Favuzzi}, {Fegan},
  {Ferrara}, {Focke}, {Fortin}, {Fukazawa}, {Funk}, {Fusco}, {Gaggero},
  {Gargano}, {Germani}, {Giglietto}, {Giordano}, {Giroletti}, {Glanzman},
  {Godfrey}, {Grove}, {Guiriec}, {Gustafsson}, {Hadasch}, {Hanabata},
  {Harding}, {Hayashida}, {Hays}, {Horan}, {Hou}, {Hughes}, {J{\'o}hannesson},
  {Johnson}, {Johnson}, {Kamae}, {Katagiri}, {Kataoka}, {Kn{\"o}dlseder},
  {Kuss}, {Lande}, {Latronico}, {Lee}, {Lemoine-Goumard}, {Longo}, {Loparco},
  {Lott}, {Lovellette}, {Lubrano}, {Mazziotta}, {McEnery}, {Michelson},
  {Mitthumsiri}, {Mizuno}, {Monte}, {Monzani}, {Morselli}, {Moskalenko},
  {Murgia}, {Naumann-Godo}, {Norris}, {Nuss}, {Ohsugi}, {Okumura}, {Omodei},
  {Orlando}, {Ormes}, {Paneque}, {Panetta}, {Parent}, {Pesce-Rollins},
  {Pierbattista}, {Piron}, {Pivato}, {Porter}, {Rain{\`o}}, {Rando}, {Razzano},
  {Razzaque}, {Reimer}, {Reimer}, {Sadrozinski}, {Sgr{\`o}}, {Siskind},
  {Spandre}, {Spinelli}, {Strong}, {Suson}, {Takahashi}, {Tanaka}, {Thayer},
  {Thayer}, {Thompson}, {Tibaldo}, {Tinivella}, {Torres}, {Tosti}, {Troja},
  {Usher}, {Vandenbroucke}, {Vasileiou}, {Vianello}, {Vitale}, {Waite}, {Wang},
  {Winer}, {Wood}, {Wood}, {Yang}, {Ziegler}, \& {Zimmer}}]{Ackermann:2012}
{Ackermann}, M., {Ajello}, M., {Atwood}, W.~B., {et~al.} 2012, \apj, 750, 3

\bibitem[{{Ackermann} {et~al.}(2013){Ackermann}, {Ajello}, {Allafort},
  {Baldini}, {Ballet}, {Barbiellini}, {Baring}, {Bastieri}, {Bechtol},
  {Bellazzini}, {Blandford}, {Bloom}, {Bonamente}, {Borgland}, {Bottacini},
  {Brandt}, {Bregeon}, {Brigida}, {Bruel}, {Buehler}, {Busetto}, {Buson},
  {Caliandro}, {Cameron}, {Caraveo}, {Casandjian}, {Cecchi}, {{\c C}elik},
  {Charles}, {Chaty}, {Chaves}, {Chekhtman}, {Cheung}, {Chiang}, {Chiaro},
  {Cillis}, {Ciprini}, {Claus}, {Cohen-Tanugi}, {Cominsky}, {Conrad}, {Corbel},
  {Cutini}, {D'Ammando}, {de Angelis}, {de Palma}, {Dermer}, {do Couto e
  Silva}, {Drell}, {Drlica-Wagner}, {Falletti}, {Favuzzi}, {Ferrara},
  {Franckowiak}, {Fukazawa}, {Funk}, {Fusco}, {Gargano}, {Germani},
  {Giglietto}, {Giommi}, {Giordano}, {Giroletti}, {Glanzman}, {Godfrey},
  {Grenier}, {Grondin}, {Grove}, {Guiriec}, {Hadasch}, {Hanabata}, {Harding},
  {Hayashida}, {Hayashi}, {Hays}, {Hewitt}, {Hill}, {Hughes}, {Jackson},
  {Jogler}, {J{\'o}hannesson}, {Johnson}, {Kamae}, {Kataoka}, {Katsuta},
  {Kn{\"o}dlseder}, {Kuss}, {Lande}, {Larsson}, {Latronico}, {Lemoine-Goumard},
  {Longo}, {Loparco}, {Lovellette}, {Lubrano}, {Madejski}, {Massaro}, {Mayer},
  {Mazziotta}, {McEnery}, {Mehault}, {Michelson}, {Mignani}, {Mitthumsiri},
  {Mizuno}, {Moiseev}, {Monzani}, {Morselli}, {Moskalenko}, {Murgia},
  {Nakamori}, {Nemmen}, {Nuss}, {Ohno}, {Ohsugi}, {Omodei}, {Orienti},
  {Orlando}, {Ormes}, {Paneque}, {Perkins}, {Pesce-Rollins}, {Piron}, {Pivato},
  {Rain{\`o}}, {Rando}, {Razzano}, {Razzaque}, {Reimer}, {Reimer}, {Ritz},
  {Romoli}, {S{\'a}nchez-Conde}, {Schulz}, {Sgr{\`o}}, {Simeon}, {Siskind},
  {Smith}, {Spandre}, {Spinelli}, {Stecker}, {Strong}, {Suson}, {Tajima},
  {Takahashi}, {Takahashi}, {Tanaka}, {Thayer}, {Thayer}, {Thompson},
  {Thorsett}, {Tibaldo}, {Tibolla}, {Tinivella}, {Troja}, {Uchiyama}, {Usher},
  {Vandenbroucke}, {Vasileiou}, {Vianello}, {Vitale}, {Waite}, {Werner},
  {Winer}, {Wood}, {Wood}, {Yamazaki}, {Yang}, \& {Zimmer}}]{Ackermann:2013}
{Ackermann}, M., {Ajello}, M., {Allafort}, A., {et~al.} 2013, Science, 339, 807

\bibitem[{{Agertz} {et~al.}(2013){Agertz}, {Kravtsov}, {Leitner}, \&
  {Gnedin}}]{Agertz:2013}
{Agertz}, O., {Kravtsov}, A.~V., {Leitner}, S.~N., \& {Gnedin}, N.~Y. 2013,
  \apj, 770, 25

\bibitem[{{Binney}(1977)}]{Binney:1977}
{Binney}, J. 1977, \apj, 215, 483

\bibitem[{{Booth} {et~al.}(2013){Booth}, {Agertz}, {Kravtsov}, \&
  {Gnedin}}]{Booth:2013}
{Booth}, C.~M., {Agertz}, O., {Kravtsov}, A.~V., \& {Gnedin}, N.~Y. 2013,
  \apjl, 777, L16

\bibitem[{{Bordoloi} {et~al.}(2011){Bordoloi}, {Lilly}, {Knobel}, {Bolzonella},
  {Kampczyk}, {Carollo}, {Iovino}, {Zucca}, {Contini}, {Kneib}, {Le Fevre},
  {Mainieri}, {Renzini}, {Scodeggio}, {Zamorani}, {Balestra}, {Bardelli},
  {Bongiorno}, {Caputi}, {Cucciati}, {de la Torre}, {de Ravel}, {Garilli},
  {Kova{\v c}}, {Lamareille}, {Le Borgne}, {Le Brun}, {Maier}, {Mignoli},
  {Pello}, {Peng}, {Perez Montero}, {Presotto}, {Scarlata}, {Silverman},
  {Tanaka}, {Tasca}, {Tresse}, {Vergani}, {Barnes}, {Cappi}, {Cimatti},
  {Coppa}, {Diener}, {Franzetti}, {Koekemoer}, {L{\'o}pez-Sanjuan},
  {McCracken}, {Moresco}, {Nair}, {Oesch}, {Pozzetti}, \&
  {Welikala}}]{Bordoloi:2011}
{Bordoloi}, R., {Lilly}, S.~J., {Knobel}, C., {et~al.} 2011, \apj, 743, 10

\bibitem[{{Boulares} \& {Cox}(1990)}]{Boulares:1990}
{Boulares}, A., \& {Cox}, D.~P. 1990, \apj, 365, 544

\bibitem[{{Brio} \& {Wu}(1988)}]{Brio:1988}
{Brio}, M., \& {Wu}, C.~C. 1988, Journal of Computational Physics, 75, 400

\bibitem[{{Bryan} {et~al.}(2014){Bryan}, {Norman}, {O'Shea}, {Abel}, {Wise},
  {Turk}, {Reynolds}, {Collins}, {Wang}, {Skillman}, {Smith}, {Harkness},
  {Bordner}, {Kim}, {Kuhlen}, {Xu}, {Goldbaum}, {Hummels}, {Kritsuk}, {Tasker},
  {Skory}, {Simpson}, {Hahn}, {Oishi}, {So}, {Zhao}, {Cen}, {Li}, \& {Enzo
  Collaboration}}]{Bryan:2014}
{Bryan}, G.~L., {Norman}, M.~L., {O'Shea}, B.~W., {et~al.} 2014, \apjs, 211, 19

\bibitem[{{Butsky} {et~al.}(2017){Butsky}, {Zrake}, {Kim}, {Yang}, \&
  {Abel}}]{Butsky:2017}
{Butsky}, I., {Zrake}, J., {Kim}, J.-h., {Yang}, H.-I., \& {Abel}, T. 2017,
  \apj, 843, 113

\bibitem[{{Cen} \& {Ostriker}(1992)}]{Cen:1992}
{Cen}, R., \& {Ostriker}, J.~P. 1992, \apjl, 399, L113

\bibitem[{{Chen} {et~al.}(2010){Chen}, {Helsby}, {Gauthier}, {Shectman},
  {Thompson}, \& {Tinker}}]{Chen:2010}
{Chen}, H.-W., {Helsby}, J.~E., {Gauthier}, J.-R., {et~al.} 2010, \apj, 714,
  1521

\bibitem[{{Cheng} {et~al.}(1994){Cheng}, {Schramm}, \& {Truran}}]{Cheng:1994}
{Cheng}, B., {Schramm}, D.~N., \& {Truran}, J.~W. 1994, \prd, 49, 5006

\bibitem[{{Christensen} {et~al.}(2014){Christensen}, {Brooks}, {Fisher},
  {Governato}, {McCleary}, {Quinn}, {Shen}, \& {Wadsley}}]{Christensen:2014}
{Christensen}, C.~R., {Brooks}, A.~M., {Fisher}, D.~B., {et~al.} 2014, \mnras,
  440, L51

\bibitem[{{Collins} {et~al.}(2010){Collins}, {Xu}, {Norman}, {Li}, \&
  {Li}}]{Collins:2010}
{Collins}, D.~C., {Xu}, H., {Norman}, M.~L., {Li}, H., \& {Li}, S. 2010, \apjs,
  186, 308

\bibitem[{{Creasey} {et~al.}(2013){Creasey}, {Theuns}, \&
  {Bower}}]{Creasey:2013}
{Creasey}, P., {Theuns}, T., \& {Bower}, R.~G. 2013, \mnras, 429, 1922

\bibitem[{{Dav{\'e}} {et~al.}(2011){Dav{\'e}}, {Oppenheimer}, \&
  {Finlator}}]{Dave:2011}
{Dav{\'e}}, R., {Oppenheimer}, B.~D., \& {Finlator}, K. 2011, \mnras, 415, 11

\bibitem[{{Dedner} {et~al.}(2002){Dedner}, {Kemm}, {Kr{\"o}ner}, {Munz},
  {Schnitzer}, \& {Wesenberg}}]{Dedner:2002}
{Dedner}, A., {Kemm}, F., {Kr{\"o}ner}, D., {et~al.} 2002, Journal of
  Computational Physics, 175, 645

\bibitem[{{Drury} \& {Falle}(1986)}]{Drury:1986}
{Drury}, L.~O., \& {Falle}, S.~A.~E.~G. 1986, \mnras, 223, 353

\bibitem[{{Durrer} \& {Neronov}(2013)}]{Durrer:2013}
{Durrer}, R., \& {Neronov}, A. 2013, \aapr, 21, 62

\bibitem[{{Ellison} {et~al.}(2010){Ellison}, {Patnaude}, {Slane}, \&
  {Raymond}}]{Ellison:2010}
{Ellison}, D.~C., {Patnaude}, D.~J., {Slane}, P., \& {Raymond}, J. 2010, \apj,
  712, 287

\bibitem[{{En{\ss}lin} {et~al.}(2007){En{\ss}lin}, {Pfrommer}, {Springel}, \&
  {Jubelgas}}]{Ensslin:2007}
{En{\ss}lin}, T.~A., {Pfrommer}, C., {Springel}, V., \& {Jubelgas}, M. 2007,
  \aap, 473, 41

\bibitem[{{Farber} {et~al.}(2017){Farber}, {Ruszkowski}, {Yang}, \&
  {Zweibel}}]{Farber:2017}
{Farber}, R., {Ruszkowski}, M., {Yang}, H.-Y.~K., \& {Zweibel}, E.~G. 2017,
  ArXiv e-prints, arXiv:1707.04579

\bibitem[{{Feldmann} {et~al.}(2013){Feldmann}, {Hooper}, \&
  {Gnedin}}]{Feldmann:2013}
{Feldmann}, R., {Hooper}, D., \& {Gnedin}, N.~Y. 2013, \apj, 763, 21

\bibitem[{{Ferland} {et~al.}(2013){Ferland}, {Porter}, {van Hoof}, {Williams},
  {Abel}, {Lykins}, {Shaw}, {Henney}, \& {Stancil}}]{Ferland:2013}
{Ferland}, G.~J., {Porter}, R.~L., {van Hoof}, P.~A.~M., {et~al.} 2013, Rev.
  Mexicana Astron. Astrofis., 49, 137

\bibitem[{{Fielding} {et~al.}(2017){Fielding}, {Quataert}, {McCourt}, \&
  {Thompson}}]{Fielding:2017}
{Fielding}, D., {Quataert}, E., {McCourt}, M., \& {Thompson}, T.~A. 2017,
  \mnras, 466, 3810

\bibitem[{{Gauthier} {et~al.}(2010){Gauthier}, {Chen}, \&
  {Tinker}}]{Gauthier:2010}
{Gauthier}, J.-R., {Chen}, H.-W., \& {Tinker}, J.~L. 2010, \apj, 716, 1263

\bibitem[{{Girichidis} {et~al.}(2016){Girichidis}, {Naab}, {Walch}, {Hanasz},
  {Mac Low}, {Ostriker}, {Gatto}, {Peters}, {W{\"u}nsch}, {Glover}, {Klessen},
  {Clark}, \& {Baczynski}}]{Girichidis:2016}
{Girichidis}, P., {Naab}, T., {Walch}, S., {et~al.} 2016, \apjl, 816, L19

\bibitem[{{Governato} {et~al.}(2012){Governato}, {Zolotov}, {Pontzen},
  {Christensen}, {Oh}, {Brooks}, {Quinn}, {Shen}, \&
  {Wadsley}}]{Governato:2012}
{Governato}, F., {Zolotov}, A., {Pontzen}, A., {et~al.} 2012, \mnras, 422, 1231

\bibitem[{{Green} {et~al.}(2012){Green}, {Froning}, {Osterman}, {Ebbets},
  {Heap}, {Leitherer}, {Linsky}, {Savage}, {Sembach}, {Shull}, {Siegmund},
  {Snow}, {Spencer}, {Stern}, {Stocke}, {Welsh}, {B{\'e}land}, {Burgh},
  {Danforth}, {France}, {Keeney}, {McPhate}, {Penton}, {Andrews},
  {Brownsberger}, {Morse}, \& {Wilkinson}}]{Green:2012}
{Green}, J.~C., {Froning}, C.~S., {Osterman}, S., {et~al.} 2012, \apj, 744, 60

\bibitem[{{Gutcke} {et~al.}(2017){Gutcke}, {Stinson}, {Macci{\`o}}, {Wang}, \&
  {Dutton}}]{Gutcke:2017}
{Gutcke}, T.~A., {Stinson}, G.~S., {Macci{\`o}}, A.~V., {Wang}, L., \&
  {Dutton}, A.~A. 2017, \mnras, 464, 2796

\bibitem[{{Haardt} \& {Madau}(2012)}]{Haardt:2012}
{Haardt}, F., \& {Madau}, P. 2012, \apj, 746, 125

\bibitem[{{Hernquist}(1990)}]{Hernquist:1990}
{Hernquist}, L. 1990, \apj, 356, 359

\bibitem[{{Hopkins} {et~al.}(2012){Hopkins}, {Quataert}, \&
  {Murray}}]{Hopkins:2012}
{Hopkins}, P.~F., {Quataert}, E., \& {Murray}, N. 2012, \mnras, 421, 3522

\bibitem[{{Hummels} \& {Bryan}(2012)}]{Hummels:2012}
{Hummels}, C.~B., \& {Bryan}, G.~L. 2012, \apj, 749, 140

\bibitem[{{Hummels} {et~al.}(2013){Hummels}, {Bryan}, {Smith}, \&
  {Turk}}]{Hummels:2013}
{Hummels}, C.~B., {Bryan}, G.~L., {Smith}, B.~D., \& {Turk}, M.~J. 2013,
  \mnras, 430, 1548

\bibitem[{{Hummels} {et~al.}(2017){Hummels}, {Smith}, \&
  {Silvia}}]{Hummels:2017}
{Hummels}, C.~B., {Smith}, B.~D., \& {Silvia}, D.~W. 2017, \apj, 847, 59

\bibitem[{{Jacob} \& {Pfrommer}(2017)}]{Jacob:2017}
{Jacob}, S., \& {Pfrommer}, C. 2017, \mnras, 467, 1478

\bibitem[{{Ji} {et~al.}(2017){Ji}, {Oh}, \& {McCourt}}]{Ji:2017}
{Ji}, S., {Oh}, S.~P., \& {McCourt}, M. 2017, ArXiv e-prints, arXiv:1710.00822

\bibitem[{{Jiang} \& {Oh}(2017)}]{Jiang:2018}
{Jiang}, Y.-F., \& {Oh}, P. 2017, ArXiv e-prints, arXiv:1712.07117

\bibitem[{{Joung} {et~al.}(2009){Joung}, {Mac Low}, \& {Bryan}}]{Joung:2009}
{Joung}, M.~R., {Mac Low}, M.-M., \& {Bryan}, G.~L. 2009, \apj, 704, 137

\bibitem[{{Jubelgas} {et~al.}(2008){Jubelgas}, {Springel}, {En{\ss}lin}, \&
  {Pfrommer}}]{Jubelgas:2008}
{Jubelgas}, M., {Springel}, V., {En{\ss}lin}, T., \& {Pfrommer}, C. 2008, \aap,
  481, 33

\bibitem[{{Jun} {et~al.}(1994){Jun}, {Clarke}, \& {Norman}}]{Jun:1994}
{Jun}, B.-I., {Clarke}, D.~A., \& {Norman}, M.~L. 1994, \apj, 429, 748

\bibitem[{{Kacprzak} {et~al.}(2010){Kacprzak}, {Churchill}, {Ceverino},
  {Steidel}, {Klypin}, \& {Murphy}}]{Kacprzak:2010}
{Kacprzak}, G.~G., {Churchill}, C.~W., {Ceverino}, D., {et~al.} 2010, \apj,
  711, 533

\bibitem[{{Keller} {et~al.}(2015){Keller}, {Wadsley}, \&
  {Couchman}}]{Keller:2015}
{Keller}, B.~W., {Wadsley}, J., \& {Couchman}, H.~M.~P. 2015, \mnras, 453, 3499

\bibitem[{{Keller} {et~al.}(2016){Keller}, {Wadsley}, \&
  {Couchman}}]{Keller:2016}
---. 2016, \mnras, 463, 1431

\bibitem[{{Kim} {et~al.}(2011){Kim}, {Wise}, {Alvarez}, \& {Abel}}]{Kim:2011}
{Kim}, J.-h., {Wise}, J.~H., {Alvarez}, M.~A., \& {Abel}, T. 2011, \apj, 738,
  54

\bibitem[{{Kim} {et~al.}(2014){Kim}, {Abel}, {Agertz}, {Bryan}, {Ceverino},
  {Christensen}, {Conroy}, {Dekel}, {Gnedin}, {Goldbaum}, {Guedes}, {Hahn},
  {Hobbs}, {Hopkins}, {Hummels}, {Iannuzzi}, {Keres}, {Klypin}, {Kravtsov},
  {Krumholz}, {Kuhlen}, {Leitner}, {Madau}, {Mayer}, {Moody}, {Nagamine},
  {Norman}, {Onorbe}, {O'Shea}, {Pillepich}, {Primack}, {Quinn}, {Read},
  {Robertson}, {Rocha}, {Rudd}, {Shen}, {Smith}, {Szalay}, {Teyssier},
  {Thompson}, {Todoroki}, {Turk}, {Wadsley}, {Wise}, {Zolotov}, \& {AGORA
  Collaboration29}}]{Kim:2014}
{Kim}, J.-h., {Abel}, T., {Agertz}, O., {et~al.} 2014, \apjs, 210, 14

\bibitem[{{Kulsrud} \& {Pearce}(1969)}]{Kulsrud:1969}
{Kulsrud}, R., \& {Pearce}, W.~P. 1969, \apj, 156, 445

\bibitem[{{Kumar} \& {Eichler}(2014)}]{Kumar:2014}
{Kumar}, R., \& {Eichler}, D. 2014, \apj, 785, 129

\bibitem[{{Kurganov} \& {Tadmor}(2000)}]{Kurganov:2000}
{Kurganov}, A., \& {Tadmor}, E. 2000, Journal of Computational Physics, 160,
  241

\bibitem[{{Marasco} {et~al.}(2015){Marasco}, {Debattista}, {Fraternali}, {van
  der Hulst}, {Wadsley}, {Quinn}, \& {Ro{\v s}kar}}]{Marasco:2015}
{Marasco}, A., {Debattista}, V.~P., {Fraternali}, F., {et~al.} 2015, \mnras,
  451, 4223

\bibitem[{{McNamara} \& {Nulsen}(2007)}]{McNamara:2007}
{McNamara}, B.~R., \& {Nulsen}, P.~E.~J. 2007, \araa, 45, 117

\bibitem[{{McQuinn} \& {Werk}(2017)}]{McQuinn:2017}
{McQuinn}, M., \& {Werk}, J.~K. 2017, ArXiv e-prints, arXiv:1703.03422

\bibitem[{{Miniati} {et~al.}(2001){Miniati}, {Jones}, {Kang}, \&
  {Ryu}}]{Miniati:2001}
{Miniati}, F., {Jones}, T.~W., {Kang}, H., \& {Ryu}, D. 2001, \apj, 562, 233

\bibitem[{{Murray} {et~al.}(2011){Murray}, {M{\'e}nard}, \&
  {Thompson}}]{Murray:2011}
{Murray}, N., {M{\'e}nard}, B., \& {Thompson}, T.~A. 2011, \apj, 735, 66

\bibitem[{{Navarro} {et~al.}(1997){Navarro}, {Frenk}, \&
  {White}}]{Navarro:1997}
{Navarro}, J.~F., {Frenk}, C.~S., \& {White}, S.~D.~M. 1997, \apj, 490, 493

\bibitem[{{Nielsen} {et~al.}(2013){Nielsen}, {Churchill}, \&
  {Kacprzak}}]{Nielsen:2013}
{Nielsen}, N.~M., {Churchill}, C.~W., \& {Kacprzak}, G.~G. 2013, \apj, 776, 115

\bibitem[{{Oppenheimer} \& {Dav{\'e}}(2006)}]{Oppenheimer:2006}
{Oppenheimer}, B.~D., \& {Dav{\'e}}, R. 2006, \mnras, 373, 1265

\bibitem[{{Oppenheimer} \& {Schaye}(2013)}]{Oppenheimer:2013}
{Oppenheimer}, B.~D., \& {Schaye}, J. 2013, \mnras, 434, 1043

\bibitem[{{Oppenheimer} {et~al.}(2018){Oppenheimer}, {Segers}, {Schaye},
  {Richings}, \& {Crain}}]{Oppenheimer:2017}
{Oppenheimer}, B.~D., {Segers}, M., {Schaye}, J., {Richings}, A.~J., \&
  {Crain}, R.~A. 2018, \mnras, 474, 4740

\bibitem[{{Pakmor} {et~al.}(2016{\natexlab{a}}){Pakmor}, {Pfrommer}, {Simpson},
  {Kannan}, \& {Springel}}]{Pakmor:2016b}
{Pakmor}, R., {Pfrommer}, C., {Simpson}, C.~M., {Kannan}, R., \& {Springel}, V.
  2016{\natexlab{a}}, \mnras, 462, 2603

\bibitem[{{Pakmor} {et~al.}(2016{\natexlab{b}}){Pakmor}, {Pfrommer}, {Simpson},
  \& {Springel}}]{Pakmor:2016a}
{Pakmor}, R., {Pfrommer}, C., {Simpson}, C.~M., \& {Springel}, V.
  2016{\natexlab{b}}, \apjl, 824, L30

\bibitem[{{Parrish} \& {Stone}(2005)}]{Parrish:2005}
{Parrish}, I.~J., \& {Stone}, J.~M. 2005, \apj, 633, 334

\bibitem[{{Peeples} {et~al.}(2014){Peeples}, {Werk}, {Tumlinson},
  {Oppenheimer}, {Prochaska}, {Katz}, \& {Weinberg}}]{Peeples:2014}
{Peeples}, M.~S., {Werk}, J.~K., {Tumlinson}, J., {et~al.} 2014, \apj, 786, 54

\bibitem[{{Pfrommer} {et~al.}(2008){Pfrommer}, {En{\ss}lin}, \&
  {Springel}}]{Pfrommer:2008}
{Pfrommer}, C., {En{\ss}lin}, T.~A., \& {Springel}, V. 2008, \mnras, 385, 1211

\bibitem[{{Pfrommer} {et~al.}(2006){Pfrommer}, {Springel}, {En{\ss}lin}, \&
  {Jubelgas}}]{Pfrommer:2006}
{Pfrommer}, C., {Springel}, V., {En{\ss}lin}, T.~A., \& {Jubelgas}, M. 2006,
  \mnras, 367, 113

\bibitem[{{Pontzen} {et~al.}(2013){Pontzen}, {Ro{\v s}kar}, {Stinson}, {Woods},
  {Reed}, {Coles}, \& {Quinn}}]{Pontzen:2013}
{Pontzen}, A., {Ro{\v s}kar}, R., {Stinson}, G.~S., {et~al.} 2013, {pynbody:
  Astrophysics Simulation Analysis for Python}, astrophysics Source Code
  Library, ascl:1305.002

\bibitem[{{Prochaska} {et~al.}(2011){Prochaska}, {Weiner}, {Chen}, {Mulchaey},
  \& {Cooksey}}]{Prochaska:2011}
{Prochaska}, J.~X., {Weiner}, B., {Chen}, H.-W., {Mulchaey}, J., \& {Cooksey},
  K. 2011, \apj, 740, 91

\bibitem[{{Ptuskin} {et~al.}(2006){Ptuskin}, {Moskalenko}, {Jones}, {Strong},
  \& {Zirakashvili}}]{Ptuskin:2006}
{Ptuskin}, V.~S., {Moskalenko}, I.~V., {Jones}, F.~C., {Strong}, A.~W., \&
  {Zirakashvili}, V.~N. 2006, \apj, 642, 902

\bibitem[{{Puchwein} \& {Springel}(2013)}]{Puchwein:2013}
{Puchwein}, E., \& {Springel}, V. 2013, \mnras, 428, 2966

\bibitem[{{Rees} \& {Ostriker}(1977)}]{Rees:1977}
{Rees}, M.~J., \& {Ostriker}, J.~P. 1977, \mnras, 179, 541

\bibitem[{{Rudie} {et~al.}(2012){Rudie}, {Steidel}, {Trainor}, {Rakic},
  {Bogosavljevi{\'c}}, {Pettini}, {Reddy}, {Shapley}, {Erb}, \&
  {Law}}]{Rudie:2012}
{Rudie}, G.~C., {Steidel}, C.~C., {Trainor}, R.~F., {et~al.} 2012, \apj, 750,
  67

\bibitem[{{Ruszkowski} {et~al.}(2017){Ruszkowski}, {Yang}, \&
  {Zweibel}}]{Ruszkowski:2017}
{Ruszkowski}, M., {Yang}, H.-Y.~K., \& {Zweibel}, E. 2017, \apj, 834, 208

\bibitem[{{Salem} \& {Bryan}(2014)}]{Salem:2014a}
{Salem}, M., \& {Bryan}, G.~L. 2014, \mnras, 437, 3312

\bibitem[{{Salem} {et~al.}(2016){Salem}, {Bryan}, \& {Corlies}}]{Salem:2016}
{Salem}, M., {Bryan}, G.~L., \& {Corlies}, L. 2016, \mnras, 456, 582

\bibitem[{{Samui} {et~al.}(2018){Samui}, {Subramanian}, \&
  {Srianand}}]{Samui:2017}
{Samui}, S., {Subramanian}, K., \& {Srianand}, R. 2018, \mnras,
  arXiv:1706.01890

\bibitem[{{Schaye} {et~al.}(2010){Schaye}, {Dalla Vecchia}, {Booth}, {Wiersma},
  {Theuns}, {Haas}, {Bertone}, {Duffy}, {McCarthy}, \& {van de
  Voort}}]{Schaye:2010}
{Schaye}, J., {Dalla Vecchia}, C., {Booth}, C.~M., {et~al.} 2010, \mnras, 402,
  1536

\bibitem[{{Sharma} \& {Nath}(2012)}]{Sharma:2012}
{Sharma}, M., \& {Nath}, B.~B. 2012, \apj, 750, 55

\bibitem[{{Sharma} {et~al.}(2009){Sharma}, {Colella}, \&
  {Martin}}]{Sharma:2009}
{Sharma}, P., {Colella}, P., \& {Martin}, D.~F. 2009, ArXiv e-prints,
  arXiv:0909.5426

\bibitem[{{Sharma} \& {Hammett}(2007)}]{Sharma:2007}
{Sharma}, P., \& {Hammett}, G.~W. 2007, Journal of Computational Physics, 227,
  123

\bibitem[{{Shu} \& {Osher}(1988)}]{Shu:1988}
{Shu}, C.-W., \& {Osher}, S. 1988, Journal of Computational Physics, 77, 439

\bibitem[{{Silk}(1977)}]{Silk:1977}
{Silk}, J. 1977, \apj, 211, 638

\bibitem[{{Simpson} {et~al.}(2016){Simpson}, {Pakmor}, {Marinacci}, {Pfrommer},
  {Springel}, {Glover}, {Clark}, \& {Smith}}]{Simpson:2016}
{Simpson}, C.~M., {Pakmor}, R., {Marinacci}, F., {et~al.} 2016, \apjl, 827, L29

\bibitem[{{Smith} {et~al.}(2017){Smith}, {Bryan}, {Glover}, {Goldbaum}, {Turk},
  {Regan}, {Wise}, {Schive}, {Abel}, {Emerick}, {O'Shea}, {Anninos}, {Hummels},
  \& {Khochfar}}]{Smith:2017}
{Smith}, B.~D., {Bryan}, G.~L., {Glover}, S.~C.~O., {et~al.} 2017, \mnras, 466,
  2217

\bibitem[{{Socrates} {et~al.}(2008){Socrates}, {Davis}, \&
  {Ramirez-Ruiz}}]{Socrates:2008}
{Socrates}, A., {Davis}, S.~W., \& {Ramirez-Ruiz}, E. 2008, \apj, 687, 202

\bibitem[{{Sod}(1978)}]{Sod:1978}
{Sod}, G.~A. 1978, Journal of Computational Physics, 27, 1

\bibitem[{{Springel} \& {Hernquist}(2003)}]{Springel:2003}
{Springel}, V., \& {Hernquist}, L. 2003, \mnras, 339, 289

\bibitem[{{Steidel} {et~al.}(2010){Steidel}, {Erb}, {Shapley}, {Pettini},
  {Reddy}, {Bogosavljevi{\'c}}, {Rudie}, \& {Rakic}}]{Steidel:2010}
{Steidel}, C.~C., {Erb}, D.~K., {Shapley}, A.~E., {et~al.} 2010, \apj, 717, 289

\bibitem[{{Stinson} {et~al.}(2006){Stinson}, {Seth}, {Katz}, {Wadsley},
  {Governato}, \& {Quinn}}]{Stinson:2006}
{Stinson}, G., {Seth}, A., {Katz}, N., {et~al.} 2006, \mnras, 373, 1074

\bibitem[{{Stinson} {et~al.}(2013){Stinson}, {Bovy}, {Rix}, {Brook}, {Ro{\v
  s}kar}, {Dalcanton}, {Macci{\`o}}, {Wadsley}, {Couchman}, \&
  {Quinn}}]{Stinson:2013}
{Stinson}, G.~S., {Bovy}, J., {Rix}, H.-W., {et~al.} 2013, \mnras, 436, 625

\bibitem[{{Stocke} {et~al.}(2013){Stocke}, {Keeney}, {Danforth}, {Shull},
  {Froning}, {Green}, {Penton}, \& {Savage}}]{Stocke:2013}
{Stocke}, J.~T., {Keeney}, B.~A., {Danforth}, C.~W., {et~al.} 2013, \apj, 763,
  148

\bibitem[{{Stone} \& {Norman}(1992)}]{Stone:1992}
{Stone}, J.~M., \& {Norman}, M.~L. 1992, \apjs, 80, 753

\bibitem[{{Strong} \& {Moskalenko}(1998)}]{Strong:1998}
{Strong}, A.~W., \& {Moskalenko}, I.~V. 1998, \apj, 509, 212

\bibitem[{{Suresh} {et~al.}(2015){Suresh}, {Bird}, {Vogelsberger}, {Genel},
  {Torrey}, {Sijacki}, {Springel}, \& {Hernquist}}]{Suresh:2015}
{Suresh}, J., {Bird}, S., {Vogelsberger}, M., {et~al.} 2015, \mnras, 448, 895

\bibitem[{{Tabatabaei} {et~al.}(2013){Tabatabaei}, {Schinnerer}, {Murphy},
  {Beck}, {Groves}, {Meidt}, {Krause}, {Rix}, {Sandstrom}, {Crocker},
  {Galametz}, {Helou}, {Wilson}, {Kennicutt}, {Calzetti}, {Draine}, {Aniano},
  {Dale}, {Dumas}, {Engelbracht}, {Gordon}, {Hinz}, {Kreckel}, {Montiel}, \&
  {Roussel}}]{Tabatabaei:2013}
{Tabatabaei}, F.~S., {Schinnerer}, E., {Murphy}, E.~J., {et~al.} 2013, \aap,
  552, A19

\bibitem[{{Tasker} \& {Bryan}(2008)}]{Tasker:2008}
{Tasker}, E.~J., \& {Bryan}, G.~L. 2008, \apj, 673, 810

\bibitem[{{Thom} {et~al.}(2012){Thom}, {Tumlinson}, {Werk}, {Prochaska},
  {Oppenheimer}, {Peeples}, {Tripp}, {Katz}, {O'Meara}, {Ford}, {Dav{\'e}},
  {Sembach}, \& {Weinberg}}]{Thom:2012}
{Thom}, C., {Tumlinson}, J., {Werk}, J.~K., {et~al.} 2012, \apjl, 758, L41

\bibitem[{{Tremmel} {et~al.}(2017){Tremmel}, {Karcher}, {Governato},
  {Volonteri}, {Quinn}, {Pontzen}, {Anderson}, \& {Bellovary}}]{Tremmel:2017}
{Tremmel}, M., {Karcher}, M., {Governato}, F., {et~al.} 2017, \mnras, 470, 1121

\bibitem[{{Tremonti} {et~al.}(2004){Tremonti}, {Heckman}, {Kauffmann},
  {Brinchmann}, {Charlot}, {White}, {Seibert}, {Peng}, {Schlegel}, {Uomoto},
  {Fukugita}, \& {Brinkmann}}]{Tremonti:2004}
{Tremonti}, C.~A., {Heckman}, T.~M., {Kauffmann}, G., {et~al.} 2004, \apj, 613,
  898

\bibitem[{{Tumlinson} {et~al.}(2011){Tumlinson}, {Werk}, {Thom}, {Meiring},
  {Prochaska}, {Tripp}, {O'Meara}, {Okrochkov}, \& {Sembach}}]{Tumlinson:2011}
{Tumlinson}, J., {Werk}, J.~K., {Thom}, C., {et~al.} 2011, \apj, 733, 111

\bibitem[{{Turk} {et~al.}(2012){Turk}, {Oishi}, {Abel}, \& {Bryan}}]{Turk:2012}
{Turk}, M.~J., {Oishi}, J.~S., {Abel}, T., \& {Bryan}, G.~L. 2012, \apj, 745,
  154

\bibitem[{{Turner} {et~al.}(2014){Turner}, {Schaye}, {Steidel}, {Rudie}, \&
  {Strom}}]{Turner:2014}
{Turner}, M.~L., {Schaye}, J., {Steidel}, C.~C., {Rudie}, G.~C., \& {Strom},
  A.~L. 2014, \mnras, 445, 794

\bibitem[{{Uhlig} {et~al.}(2012){Uhlig}, {Pfrommer}, {Sharma}, {Nath},
  {En{\ss}lin}, \& {Springel}}]{Uhlig:2012}
{Uhlig}, M., {Pfrommer}, C., {Sharma}, M., {et~al.} 2012, \mnras, 423, 2374

\bibitem[{{van Leer}(1977)}]{VanLeer:1977}
{van Leer}, B. 1977, Journal of Computational Physics, 23, 276

\bibitem[{{Vazza} {et~al.}(2012){Vazza}, {Br{\"u}ggen}, {Gheller}, \&
  {Brunetti}}]{Vazza:2012}
{Vazza}, F., {Br{\"u}ggen}, M., {Gheller}, C., \& {Brunetti}, G. 2012, \mnras,
  421, 3375

\bibitem[{{Wang} \& {Abel}(2009)}]{Wang:2009}
{Wang}, P., \& {Abel}, T. 2009, \apj, 696, 96

\bibitem[{{Wang} {et~al.}(2008){Wang}, {Abel}, \& {Zhang}}]{Wang:2008}
{Wang}, P., {Abel}, T., \& {Zhang}, W. 2008, \apjs, 176, 467

\bibitem[{{Wefel}(1987)}]{Wefel:1987}
{Wefel}, J.~P. 1987, {Origin and transport of high energy particles in the
  galaxy}, Tech. rep.

\bibitem[{{Werk} {et~al.}(2013){Werk}, {Prochaska}, {Thom}, {Tumlinson},
  {Tripp}, {O'Meara}, \& {Peeples}}]{Werk:2013}
{Werk}, J.~K., {Prochaska}, J.~X., {Thom}, C., {et~al.} 2013, \apjs, 204, 17

\bibitem[{{Werk} {et~al.}(2014){Werk}, {Prochaska}, {Tumlinson}, {Peeples},
  {Tripp}, {Fox}, {Lehner}, {Thom}, {O'Meara}, {Ford}, {Bordoloi}, {Katz},
  {Tejos}, {Oppenheimer}, {Dav{\'e}}, \& {Weinberg}}]{Werk:2014}
{Werk}, J.~K., {Prochaska}, J.~X., {Tumlinson}, J., {et~al.} 2014, \apj, 792, 8

\bibitem[{{Wiener} {et~al.}(2017){Wiener}, {Pfrommer}, \& {Peng
  Oh}}]{Wiener:2017}
{Wiener}, J., {Pfrommer}, C., \& {Peng Oh}, S. 2017, \mnras, 467, 906

\bibitem[{{Wise} {et~al.}(2012){Wise}, {Abel}, {Turk}, {Norman}, \&
  {Smith}}]{Wise:2012}
{Wise}, J.~H., {Abel}, T., {Turk}, M.~J., {Norman}, M.~L., \& {Smith}, B.~D.
  2012, \mnras, 427, 311

\bibitem[{{Zhang} {et~al.}(2016){Zhang}, {Zaritsky}, {Zhu}, {M{\'e}nard}, \&
  {Hogg}}]{Zhang:2016}
{Zhang}, H., {Zaritsky}, D., {Zhu}, G., {M{\'e}nard}, B., \& {Hogg}, D.~W.
  2016, \apj, 833, 276

\bibitem[{{Zweibel}(2017)}]{Zweibel:2017}
{Zweibel}, E.~G. 2017, Physics of Plasmas, 24, 055402

\end{thebibliography}

\appendix
\section{Testing the Newly Implemented CR Physics}
In \S \ref{sec:crenzo}, we described the implementation of CR physics into
the Riemann solvers in Enzo. Here, we demonstrate the
performance of our implemented CR advection, anisotropic diffusion, and streaming. 
For tests of the isotropic diffusion module, refer to \cite{Salem:2014a}.

\subsection{Modified Sod Shock-tube}
The Sod shock-tube, first described in \citet{Sod:1978} is used 
to test the behavior of gas in the presence of a strong shocks in numerical simulations. 
Because the original Sod shock-tube doesn't include CRs, we use a modified version first described 
in \citet{Pfrommer:2006}. The initial conditions are described in Table \ref{table:sod} below. 

In Figure \ref{fig:shocktube_compare} we test our implementation of the non-diffusive 
($\kappa_{\varepsilon} = 0.0$) two-fluid CR model and compare it to the 
existing implementation in the {\sc Zeus} hydro scheme in {\sc Enzo} \citep{Salem:2014a}. 
From left to right, the panels show the density, velocity, and pressures of this
shock-tube after t = 0.31 code units of evolution. Where applicable, the analytic 
solution is depicted as a black dashed line. Compared to \citet{Salem:2014a}, our 
implementation has a sharper velocity profile at $x \simeq 250$ cm, but slightly 
under-predicts the density value around $x = 400$ cm. 
Overall, the two implementations of CR advection agree well with each other and with the 
analytic solution. 

\begin{table}[ht]
\centering
\caption{Initial conditions for the modified Sod Shock-tube}
\label{table:sod}
\begin{tabular}{lllll}
\hline
      & $\rho$ & $P_{g}$          & $\varepsilon_{c}$ & $v$ \\ \hline
Left  & 1.0    & $2/3 \times 10^5$ & $4.0\times10^5$   & 0.0 \\
Right & 0.2    & 267.2             & 801.6             & 0.0 \\ \hline
\end{tabular}
\footnote{The initial density, thermal pressure, CR energy density, and velocity 
for the modified Sod shocktube. The left and right regions of the shocktube are
defined as $(0 < x < 250 \mathrm{\ cm})$ and $(250 < x < 500 \mathrm{\ cm})$ respectively.}
\end{table}
\begin{table}[ht]
\centering
\caption{Modified Brio-Wu Shock-tube}
\label{tab:bw_shock}
\begin{tabular}{l|ccccccc}
\hline
      & $\rho$ & $P_{g}$ & $\varepsilon_{c}$ & $\mathrm{v_x}$ & $\mathrm{v_y}$ & $B_x$ & $B_y$ \\ \hline
Left  & 1.0    & 0.6    & 1.2 & 0.0      & 0.0 & 1.0   & 1.0   \\
Right & 0.125    & 0.06    & 0.12 & 0.0    & 0.0 & 1.0   & -1.0  \\ \hline
\end{tabular}
\footnote{The initial density, thermal pressure, CR energy density, velocity and 
magnetic field strengths for the modified Brio-Wu shocktube. The left and right 
regions of the shocktube are defined by $(-0.5 < x < 0)$ and $(0 < x < 0.5)$ respectively.}
\end{table}

\begin{figure}[ht]
\begin{center}
\includegraphics[width=\textwidth]{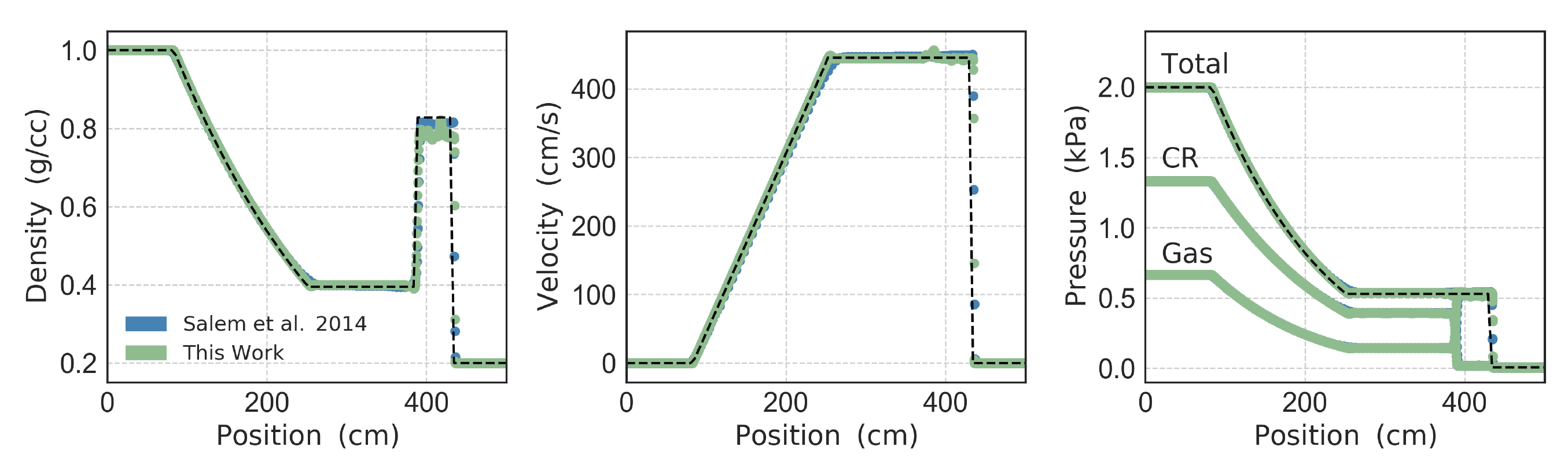}
\caption{\label{fig:shocktube_compare} The density, velocity, and pressure 
(thermal, CR, and total pressures) as a function of position in a 1D simulation
of the modified Sod shock-tube. CR pressure compliments thermal pressure while
the total pressure remained unchanged in the rarefaction wave.}
\end{center}
\end{figure}
\begin{figure}[ht]
\begin{center}
\includegraphics[width=\textwidth]{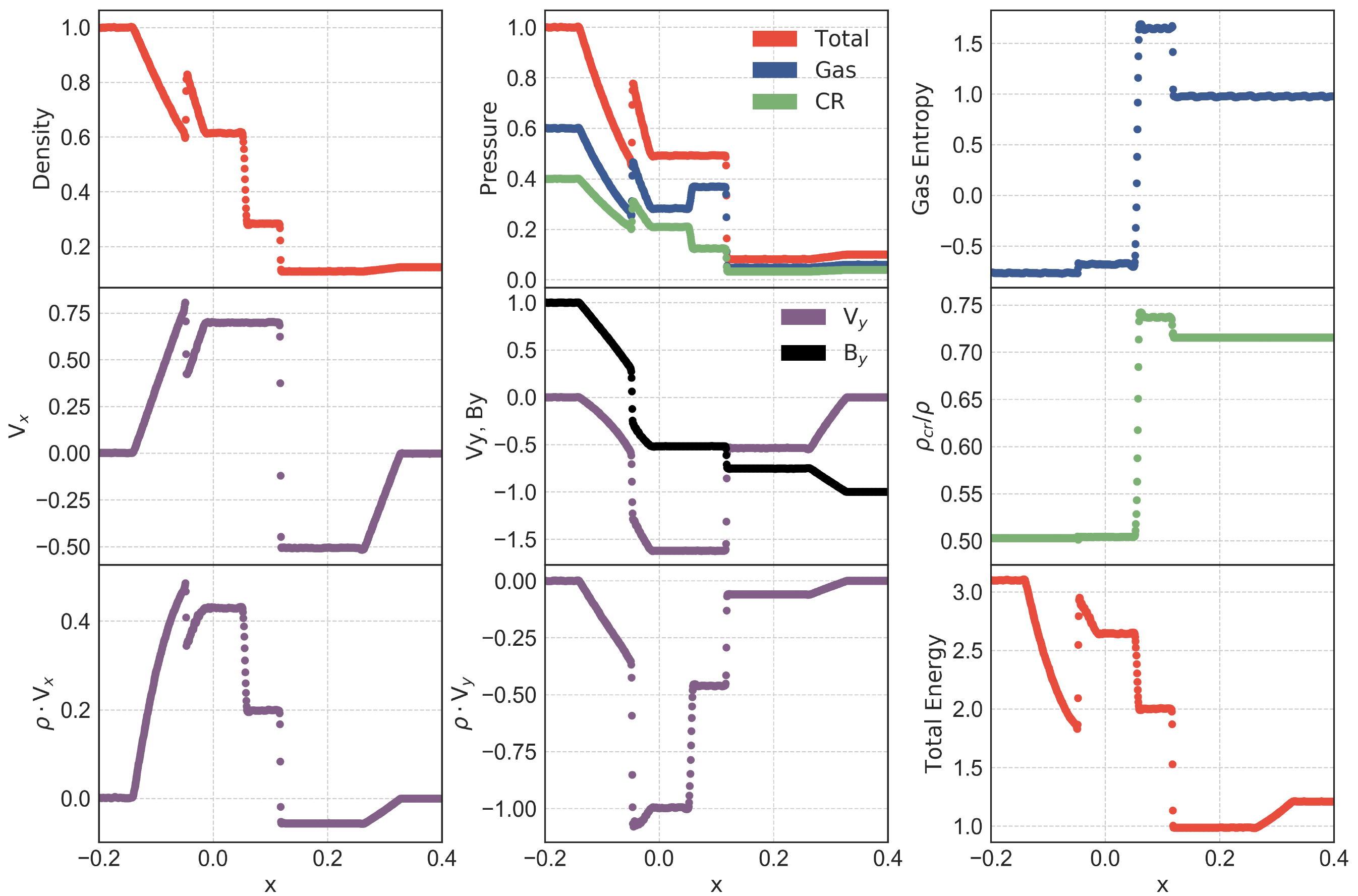}
\caption{\label{fig:bw_shock} Various parameters of the modified Brio-Wu MHD shock-tube as a function of one dimensional position. 
{\bf Top row:} density, CR, thermal, and total pressure, and thermal gas entropy. 
{\bf Middle row:} x-component of the velocity, y-component of the velocity and magnetic field, ratio of CR density to thermal density. 
{\bf Bottom row:} the density-weighted x and y components of the velocity, and the total energy. 
Thermal gas is influenced by the pressure of both CRs and magnetic field lines. }
\end{center}
\end{figure}

\subsection{Brio Wu Shock-tube}

The Brio-Wu shock-tube \cite{Brio:1988} tests the advection properties of gas interactions with magnetic fields
in the presence of an extreme shock. In the modified Brio-Wu shock-tube, we include CR pressure and evolve the
system for 0.2 code units. The initial conditions of the gas are described in Table \ref{tab:bw_shock}. 
\subsection{Anisotropic Diffusion}
\begin{figure}
\begin{center}
\includegraphics[width=\textwidth]{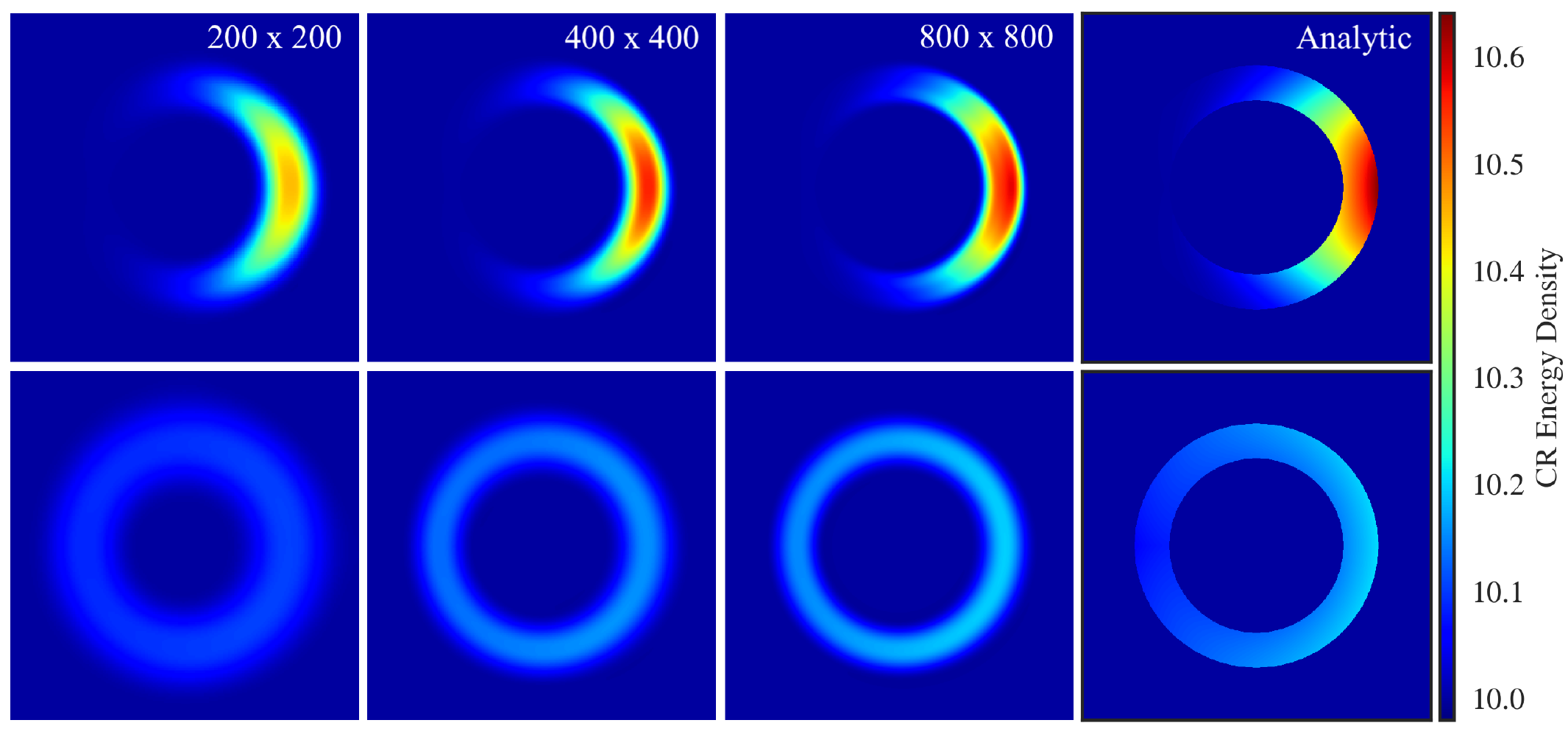}
\caption{\label{fig:pakmor}
A 2D test of anisotropic diffusion of CR energy density along circular magnetic field lines at an early time (t = 10; top row) and a late time (t = 100; bottom row).
 The columns are ordered from left to right by increasing resolution, ending with the analytic solution. With anisotropic diffusion, CRs are 
confined to propagate solely along magnetic field lines. Increasing resolution decreases the amount of diffusion perpendicular to the magnetic field. }
\end{center}
\end{figure}

In the anisotropic diffusion approximation, CRs propagate down their gradient, along 
magnetic field lines. In Figure \ref{fig:pakmor}, we test our implementation with
the  anisotropic ring problem initially described by 
\citet{Parrish:2005, Sharma:2007} and explored in detail in \citet{Pakmor:2016b}. The experiment
uses a uniform density box in with no gas (or a gas that is fixed in space such that there is no
CR advection) on a domain of $[-1,1]^2$. 

Following the setup in \citet{Pakmor:2016b}, we set initial conditions for the CR energy density to be
\begin{equation}
\epsilon_{c}(x,y) = \begin{cases}
	12\quad \mathrm{if}\enspace 0.5 < r < 0.7\enspace \mathrm{and}\enspace |\phi| < \frac{\pi}{12} \\
    10 \quad \mathrm{else},\\
\end{cases}
\end{equation}
where the radial coordinate $r = \sqrt{x^2 + y^2}$, $\phi = \mathrm{atan}2(y,x)$, and 
the diffusion coefficient is set as $\kappa_{c}= 0.01$.

In Figure \ref{fig:pakmor}, we compare the performance of our anisotropic diffusion 
implementation against the analytic solution given by
\begin{equation}
\epsilon_{c}(x,y) = \begin{cases}
10 + \mathrm{erfc}\big[\big(\phi + \frac{\pi}{12}\big)\frac{r}{D}\big] -  \mathrm{erfc}\big[\big(\phi - \frac{\pi}{12}\big)\frac{r}{D}\big]
\end{cases}
\end{equation}
where $D = \sqrt{4\kappa_{\epsilon}t}$. $\epsilon_{c}(x,y) = 10$ everywhere else. 

After evolving this simulation for 10 code time units, we see the initial wedge of uniform CR energy density moving down 
its gradient around the magnetic fields lines. The lower resolution runs have lower peak values of CR energy density and
show signs of CR transport perpendicular to the magnetic field direction. Both of these properties improve with increased resolution. 
The best-resolved simulation (800 x 800, fixed grid) matches the analytic solution well. 

At a later time (t = 100 code units), the CR energy density has reached the opposite side of the circle traced by the
toroidal magnetic field lines. The energy density approaches a steady state and is nearly evenly distributed
around the annulus. In low-resolution runs, the average final CR energy is lower, due to conservation of CR energy in the thicker rings. 

\subsection{Streaming}
\begin{figure}
\begin{center}
\includegraphics[width=0.5\textwidth]{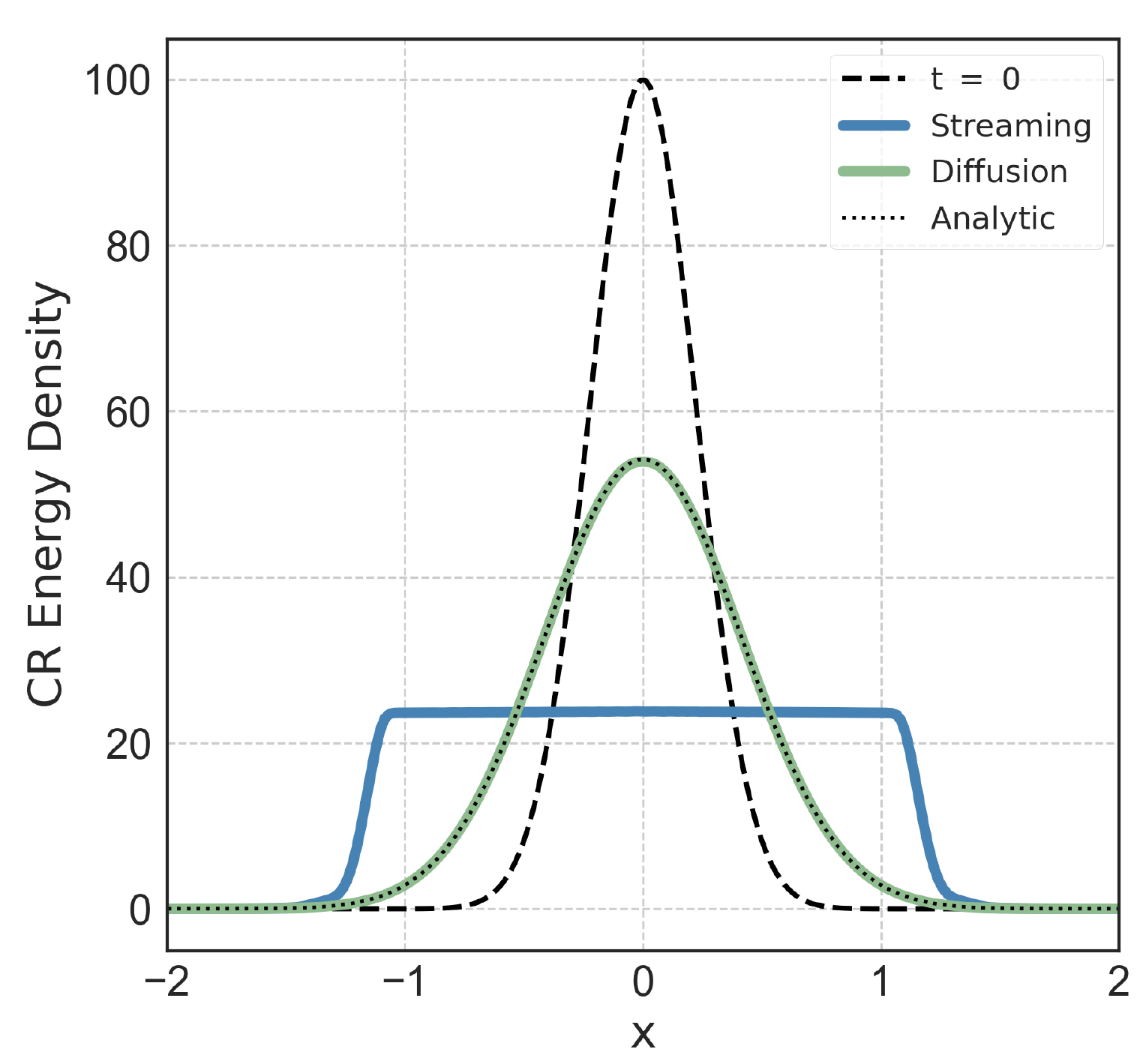}
\caption{\label{fig:streaming}
A comparison of the time evolution of a 1D Gaussian profile with CR
streaming (green) and diffusion (blue). A magnetic field of strength 
0.25 code units lies in the x direction. The analytic solution for the 
CR diffusion profile is overplotted as a black dotted line. Although no analytic
solution exists for the CR streaming case, our results qualitatively match 
previous studies. }
\end{center}
\end{figure}

We test the behavior of our CR streaming implementation with a 1D 
simulation of an initial Gaussian profile of CR energy density (see Figure \ref{fig:streaming}). 
The initial CR energy density profile is set by
\begin{equation}
\varepsilon = \varepsilon_0 e^{-x^2/2D},
\end{equation}
where $x$ is the spatial coordinate. In our example, we chose constants $\varepsilon_0 = 100$ and
$ D = 0.05$. We include a magnetic field in the $\hat{x}$ direction of a strength of 0.25 in code units.
We isolate the effects of CR streaming and diffusion by fixing the gas so that no advection is taking place.

For CR diffusion, the analytic solution for the evolution of the CR energy 
density over time is given by 
\begin{equation}
\varepsilon = \varepsilon_0 \sqrt{\frac{D}{D + 2\kappa_{\varepsilon}t}}
\mathrm{exp}\bigg(\frac{-x^2}{2(D + 2\kappa_{\varepsilon}t)}\bigg).
\end{equation}

The CR diffusion implementation follows the analytic solution well. Although there is no 
analytic solution for the CR streaming case, our results are
qualitatively comparable to 
previous studies (e.g., \cite{Uhlig:2012, Wiener:2017}).

\end{document}